\documentclass[a4paper,fleqn]{cas-sc}

\usepackage[numbers, sort&compress]{natbib}

\usepackage[utf8]{inputenc}
\usepackage{graphicx}
\usepackage{moreverb,url}
\usepackage{makecell, multirow}
\usepackage{todonotes}
\usepackage{caption}
\usepackage{subcaption}
\usepackage{booktabs}
\usepackage{siunitx}
\usepackage{floatrow}
\usepackage{ulem}
\usepackage{color}

\usepackage{bm}

\def\tsc#1{\csdef{#1}{\textsc{\lowercase{#1}}\xspace}}
\tsc{WGM}
\tsc{QE}

\begin{document}
\let\WriteBookmarks\relax
\def\floatpagepagefraction{1}
\def\textpagefraction{.001}

\shorttitle{Experimental determination of GUW dispersion diagrams in FML}    

\shortauthors{T. Barth et~al.}  

\title [mode = title]{Experimental determination of dispersion diagrams over large frequency ranges for guided ultrasonic waves in fiber metal laminates} 

\author[a]{Tilmann Barth}[orcid=0000-0002-0667-0609]
\cormark[1]
\cortext[1]{Corresponding author}
\ead{barth@hsu-hh.de}
\credit{Conceptualization, Methodology, Investigation, Data Curation, Writing - Original Draft, Writing - Review and Editing, Visualization}

\author[b]{Johannes Wiedemann}[orcid=0000-0003-2040-0143]
\credit{Conceptualization, Methodology, Investigation, Data Curation, Writing - Original Draft, Writing - Review and Editing, Visualization}

\author[b]{Thomas Roloff}[orcid=0000-0002-8834-710X]
\credit{Conceptualization, Methodology, Investigation, Data Curation, Writing - Original Draft, Writing - Review and Editing, Visualization}

\author[b]{Tim Behrens}[orcid=0000-0002-4493-445X]
\credit{Investigation, Data Curation}

\author[a]{Natalie Rauter}[orcid=0000-0003-1704-1426]
\credit{Conceptualization, Methodology, Investigation, Writing - Original Draft, Writing - Review and Editing, Visualization, Funding acquisition, Project administration}

\author[b,c]{Christian H\"{u}hne}[orcid=0000-0002-2218-1223]
\credit{Conceptualization, Writing - Review and Editing, Funding acquisition, Project administration}

\author[b]{Michael Sinapius}[orcid=0000-0002-1873-9140]
\credit{Conceptualization, Writing - Review and Editing, Funding acquisition, Project administration}

\author[a]{Rolf Lammering}[orcid=0000-0002-0867-1859]
\credit{Conceptualization, Writing - Review and Editing, Funding acquisition, Project administration}



\affiliation[a]{organization={Helmut-Schmidt-University / University of the Federal
Armed Forces Hamburg, Institute of Mechanics},
            addressline={Holstenhofweg 85}, 
            city={Hamburg},
            postcode={22043}, 
            country={Germany}}

\affiliation[b]{organization={TU Braunschweig, Institute of Mechanics and Adaptronics},
            addressline={Langer Kamp 6}, 
            city={Braunschweig},
            postcode={38106}, 
            country={Germany}}
            
\affiliation[c]{organization={DLR, Institute of Composite Structures and Adaptive Systems},
            addressline={Lilienthalplatz 7}, 
            city={Braunschweig},
            postcode={38108},
            country={Germany}}

\begin{abstract}
Fiber metal laminates (FML) are of high interest for lightweight structures as they combine the advantageous material properties of metals and fiber-reinforced polymers. However, low-velocity impacts can lead to complex internal damage. Therefore, structural health monitoring with guided ultrasonic waves (GUW) is a methodology to identify such damage.
Numerical simulations form the basis for corresponding investigations, but experimental validation of dispersion diagrams over a wide frequency range is hardly found in the literature. 
In this work the dispersive relation of GUWs is experimentally determined for an FML made of carbon fiber-reinforced polymer and steel. For this purpose, multi-frequency excitation signals are used to generate GUWs and the resulting wave field is measured via laser scanning vibrometry. The data are processed by means of a non-uniform discrete 2d Fourier transform and analyzed in the frequency-wavenumber domain. The experimental data are in excellent agreement with data from a numerical solution of the analytical framework. In conclusion, this work presents a highly automatable method to experimentally determine dispersion diagrams of GUWs in FML over large frequency ranges with high accuracy.
\end{abstract}


\begin{highlights}
\item GUW in FML made of carbon fiber-reinforced polymer (CFRP) and steel
\item Transfer of experimental method from isotropic to highly anisotropic materials (FML)
\item Experimental dispersion data for large frequency ranges in FML with high accuracy
\item Good agreement between experiment and numerical solutions from analytical framework
\end{highlights}


\maketitle

\section{Introduction}
Fiber metal laminates (FML) are advanced material systems for aeronautic applications since they show superior performance over aerospace-grade aluminum and higher ductility compared to fiber-reinforced polymers (FRP)~\cite{Vlot.2001, Alderliesten.2017}. However, due to their layered structure, low-velocity impacts can create internal damage in the material that is barely visible from the outside~\cite{Chai.2014, Moriniere.2012, Bienias.2014}. Therefore, the approach of using sensors to assess the current health state of the component at any time during the life cycle is particularly interesting for FML applications~\cite{Guemes.2006, Yan.2017}. 
\par
To detect damage in thin-walled components through structural health monitoring (SHM) systems, the use of guided ultrasonic waves (GUW) has been of major interest in research~\cite{Lammering.2017, Yuan.2016, Giurgiutiu.2008}. GUW can travel over large distances in thin solid structures without dissipating much energy and show interactions with inhomogeneities such as damage~\cite{Giurgiutiu.2008} and are therefore particularly well-suited for monitoring large-scale structures.
\par
FML specimens exhibit strong, abrupt changes in material properties across the specimen thickness. Therefore, the question arises whether the GUW's occurring in these material systems can be calculated using the methods developed for isotropic waveguides and extended for fiber composites, cf.~\cite{Lamb.1917,rose.2014}.
A first indication that the respective methods can be applied is shown by Pant et al.~\cite{Pant.2014}. Therein a 3d linear elasticity model is developed to determine the dispersion diagrams and displacement fields of GUWs in monoclinic and higher-order symmetric materials such as FML. To validate the performance of the method, GLARE3-3/4 is investigated numerically and experimentally. Using a combination of one actuator and two sensors, the group and phase velocities for the fundamental wave modes are extracted. The setup allowed an experimental validation of the numerical results for the fundamental modes. However, the $A_0$ group velocity was validated in another frequency range than the $S_0$ phase velocity. 
Muc et al.~\cite{Muc.2021} address the numerical determination of dispersion diagrams in FML using the stiffness matrix method and FEM. The study covers laminates made of aluminum and carbon fiber-reinforced polymer (CFRP) as well as aluminum and glass fiber-reinforced polymer (GFRP). The numerical results for the group velocity are compared to experimental results, using two distinct sensors to determine the time-of-flight of a single $A_0$-mode wave package at \SI{100}{\kilo\hertz}.
Gao et al.~\cite{Gao.2019} are using a combination of the state-vector formalism and the Legendre polynomials to simulate the GUW propagation in multi-layered anisotropic composite laminates. Apart from isotropic, quasi-isotropic, and anisotropic materials, the aluminum/GFRP combination standard GLARE3-3/2 is considered. The results show a good fit with the global matrix method over a broad frequency range. However, the results are not validated experimentally.
Mikhaylenko et al.~\cite{Mik.2022} investigate the displacement fields for such waves in FML made of CFRP and steel numerically, by comparing numerical solutions based on the analytical framework and on FE simulations. A good agreement could be found, but the results are also not validated experimentally.
\par
Beyond that, only little literature can be found that addresses the fundamentals of wave propagation in FMLs and especially literature targeting experimental investigations is rare. 
Maghsoodi et al.~\cite{Maghsoodi.2016} focus on damage detection methods with Lamb waves using phased arrays in FML numerically. The validation is done using ABAQUS and the transfer matrix method at a single frequency of \SI{100}{\kilo\hertz}. Tai et al.~\cite{Tai.2020} are also modeling the GUW propagation in FML numerically to investigate the effect of defects. An experimental setup is used to determine the group velocity for validating the numerical model. However, only a small frequency range of 50-\SI{250}{\kilo\hertz} with a large step size of \SI{25}{\kilo\hertz} is investigated for the $A_0$-mode.
In addition, investigations on structures with metal and FRP interfaces exist. Attar et al.~\cite{Attar.2020} perform dispersion curve measurements on the adhesive bond between aluminum and CFRP using an impulse excitation for broadband Lamb wave measurements. The experimental results match with the simulations. 
LeCrom et al.~\cite{LeCrom.2010} inspect the bonding layer between an aluminum structure and an applied CFRP patch using a laser scanning vibrometer (LSV). They determine dispersion curves for $SH$-waves in an isotropic structure with an anisotropic patch. However, these investigations focus on the characterization of bonding interfaces between those two materials with an additional bonding layer (0.2-\SI{0.5}{\milli \meter}) out of pure epoxy resin. The different layers of FMLs as considered within the scope of this work, on the other hand, are bonded during a one-step manufacturing process without any additional adhesive. Hence, an explicit bonding layer is not found in FMLs and the results presented in~\cite{Attar.2020,LeCrom.2010}  are not directly transferable to the material in this work.
\par
The state of the research shows that there is a significant need for accurate experimental GUW dispersion measurements in FML coupled with numerical investigations. Therefore, this publication aims at closing this gap by providing a highly automatable method and consequently experimentally determined dispersion diagrams for FML over a large frequency range with a high resolution. Therefore, a method originally proposed and used  by~\cite{Cawley.1991,Hora.2012,Su.2009}, which also gave very good results in the authors' work for isotropic materials~\cite{Barth.ExpM.2022}, is extended to FML. It is based on the use of 2d Fourier transformations to evaluate the measurement data of the GUWs. To do so, the structural velocity on the specimen's surface is measured with the use of an LSV. Time signals are recorded at points along a measuring path in the direction of the wave propagation to generate data as a function of time and spatial coordinates. In this work the data are evaluated by means of a non-uniform 2d discrete Fourier transform (2d-DFT)~\cite{bracewell.fourier.1986}. The method is specially tuned to increase the accuracy of the measurements as much as possible by the use of multi-frequency excitation signals instead of impulse excitations, while still enabling high automation and thus the possibility to cover large frequency ranges.
\par
The measurements are performed from three different points of view. Firstly, they are intended to provide experimental dispersion relations for FML structures over large frequency ranges to fill a gap in the existing literature. Secondly, to demonstrate the applicability of the method used for non-isotropic and inhomogeneous materials, which exhibit abrupt changes in material properties over thickness. For this purpose, repeated measurements with different measurement setups at different locations are reviewed regarding their reproducibility. Thirdly, the applicability of the presented method is compared for strip specimens with a surface area of \SI{110}{} $\times$ \SI{490}{\milli\metre} and square plate specimens with an edge length of \SI{500}{\milli\metre}. This is done in preparation for follow-up work, as strip specimens enable a cost-effective alternative to plate specimens and fulfill the boundary conditions of future investigations.
\par
The work presented here is structured as follows: After a short introduction to FMLs, the basics for the analytical description of waves in laminate structures are discussed. This is followed by the description of the experimental setups as well as the definition of the materials and specimens used in this work. Subsequently, the data generation, as well as the postprocessing, is discussed in detail. To show the performance of the presented experimental approach, several different investigations are addressed in the results section. First, the reproducibility of the measurements is discussed, followed by a comparison of different experimental setups at different locations. Furthermore, the influence of specimen size is evaluated. Finally, the measured dispersion relations are compared to numerically derived values of the analytical framework.

\section{Background}
\label{sec:theoretical_background}

\subsection{Fiber metal laminates}
FML aims at combining the advantages of metals, such as ductility, with the advantages of FRP, such as high specific strength and stiffness~\cite{Sinmazcelik.2011, Vlot.2001}. FML consist of alternating layers of thin fiber plies and metal sheets, which results in the ability to stop or bridge cracks. Thereby, a high damage tolerance is achieved~\cite{Chai.2014, Alderliesten.2003}. Moreover, advantages can be taken from the combination of the two materials, e.g. in case of impact loading~\cite{Boose.2020} or load-bearing applications~\cite{Fink.2010,Petersen.2017}, as well as for function integration~\cite{Pototzky.2019} and structure robustness~\cite{During.2020}. Despite these advantages, low-velocity impact damage can lead to internal failure of the laminate, e.g. delaminations,  which will be barely visible from the outside~\cite{Chai.2014, Moriniere.2012}. This promotes the use of SHM methods that are also sensitive to internal defects~\cite{Lammering.2018}.

\begin{figure}
\begin{floatrow}
\ffigbox[0.99\linewidth]
{\includegraphics[width=\linewidth]{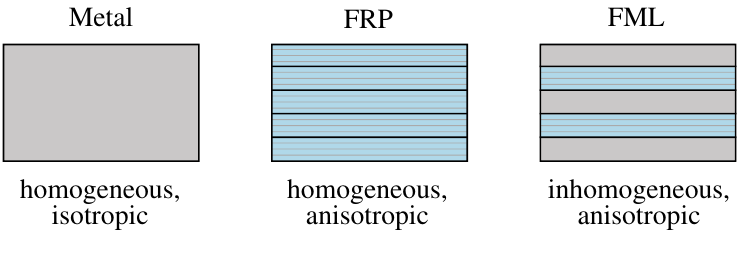}}
{\caption{FMLs compared to other structural materials in terms of homogeneity of the material and direction dependent behavior.}
\label{fig:material_classification}}
\end{floatrow}
\end{figure}

The combination of metal and FRP in a layered composite results in a material system that is highly inhomogeneous and behaves strongly anisotropic~\cite{Abouhamzeh.2016, Alderliesten.2017}. Figure~\ref{fig:material_classification} illustrates the subdivision of the different material systems according to their internal structure and directional behavior. The layered structure of the FML with its high impedance differences and interfaces between the two materials in thickness direction is assumed to have a significant influence on the propagation of the GUW. The focus in this work will be on an FML consisting of CFRP and steel.

\subsection{Waves in laminate structures}
\label{sec:waves_in_laminate_structures}
To describe the propagation of GUW in layered structures, the equation of motion is solved with respect to stress-free boundary conditions at the surfaces of thin-walled structures. The fundamental relations that are used to derive the equation of motion are the balance of momentum
\begin{equation}
\operatorname{div}\boldsymbol{\sigma} + \rho_0\mathbf{b} = \rho_0 \ddot{\mathbf{u}},
\end{equation}
the linear Green-Lagrange strain tensor 
\begin{equation}
\mathbf{E} = \frac{1}{2}\left(\operatorname{grad}^{\text{T}}\mathbf{u} + \operatorname{grad}\mathbf{u} \right),
\end{equation}
and Hooke's law
\begin{equation}
\boldsymbol{\sigma} = \bm{\mathbb{C}} : \mathbf{E}.
\end{equation}
Here, $\rho_0$ is the density, $\mathbf{b}$ the body force density, $\mathbf{u}$ the displacement field, $\boldsymbol{\sigma}$ the Cauchy stress tensor, and $\bm{\mathbb{C}}$ gives the stiffness tensor. Introducing Hooke's law in combination with the linear Green-Lagrange strain tensor into the balance of momentum and neglecting the volume forces leads to the well-known equation of motion, which is given for anisotropic materials by
\begin{equation}
\operatorname{div}\left( \bm{\mathbb{C}} : \operatorname{grad}\mathbf{u} \right) - \rho_0 \ddot{\mathbf{u}} = \mathbf{0}.
\end{equation}

Solving these differential equations for stress-free boundary conditions by using the approach
\begin{equation}
\mathbf{u} = A \mathbf{p} \operatorname{e}^{\mathrm{i}\left( \mathbf{k}\cdot\mathbf{x} - \omega t \right)}, 
\end{equation}
with the polarization vector $\mathbf{p}$, the circular wavenumber vector $\mathbf{k}$, the circular frequency $\omega$, the amplitude $A$, the spatial coordinates $\mathbf{x}$ and the time $t$ leads to the Christoffel equation in thin-walled structures for a single material layer \cite{Rokhlin.2002b}
\begin{equation}
\mathrm{i}k
\begin{bmatrix}
D_{ij}^{+} & D_{il}^{-}H_{jl} \\
D_{il}^{+}H_{jl} & D_{ij}^{-} \\
\end{bmatrix}
\begin{bmatrix}
 A_{l}^{+} \\
 A_{l}^{-} \\
\end{bmatrix}
\mathrm{e}^{\mathrm{i}\left(kx_1-\omega t\right)}
= 
\begin{bmatrix}
0 \\
0 \\
\end{bmatrix}
\label{eqn:Spannungsgls}
\end{equation}  
with
\begin{equation}
D_{ij}^{\pm} = 
\begin{bmatrix}
\left(d_{i}^{\pm}\right)_{1} &
\left(d_{i}^{\pm}\right)_{2} & 
\left(d_{i}^{\pm}\right)_{3} \\
\end{bmatrix},   
\qquad
H_{ij} = 
\begin{bmatrix}
\mathrm{e}^{\mathrm{i}k\alpha_{1}h} & 0 & 0 \\
0 & \mathrm{e}^{\mathrm{i}k\alpha_{2}h} & 0\\ 
0 & 0 & \mathrm{e}^{\mathrm{i}k\alpha_{3}h}\\
\end{bmatrix}
\qquad
\text{and}
\qquad
A_{j}^{\pm} = 
\begin{bmatrix}
A_{1}^{\pm} &
A_{2}^{\pm} & 
A_{3}^{\pm} \\
\end{bmatrix}^\mathrm{T}.
\label{eqn:AMatrix} 
\end{equation}
Here, $h$ is the plate thickness, $\alpha_i$ are the eigenvalues of the Christoffel equation, and $d_{i}^{\pm}$ can be calculated from
\begin{equation}
 \left(d_i^{\pm} \right)_n =  C_{i3kl} n_{l}^{\pm}\left(p_k\right)_n^{\pm}.
\label{eqn:Spannungsfaktor}
\end{equation}
To be able to derive dispersion diagrams for laminates, information about the layered structure must be included into the dispersion relation. A comprehensive overview of methods for dispersion diagrams in layered structures is provided in Lowe \cite{Lowe.1995}. Two common approaches are the transfer-matrix method \cite{Haskell.1990,Nayfeh.1991, Nayfeh.1995} and the global matrix method \cite{L.Knophoff.1964}. The major drawback of the transfer-matrix method is its instability for large frequency-thickness pairs. This was later solved by Kausel~\cite{Kausel.1986}, Wang~\cite{Wang.2001}, and Rokhlin~\cite{Rokhlin.2002, Rokhlin.2002b}. In contrast to this, the global matrix method is always stable and is used here to formulate the analytical framework that can be solved numerically to determine dispersion diagrams for the propagation of GUW in FML.
\par
Within the global matrix method, the presented dispersion relations for single layers are connected by introducing a continuity condition for the displacements and stresses at the layer interfaces. Therefore, in accordance with Equation~(\ref{eqn:Spannungsgls}), the displacements and out-of-plane stress components are collected in one system of equations. Following the matrix definitions in Eq.~(\ref{eqn:AMatrix}) and adding the displacement components leads to
\begin{equation}
\begin{bmatrix}
u_{1} \\
u_{2} \\
u_{3} \\
\sigma_{13}^{*}\\
\sigma_{23}^{*}\\
\sigma_{33}^{*}
\end{bmatrix}
=
\begin{bmatrix}
p_{11}^+ & p_{11}^- &  p_{12}^+ &  p_{12}^- & p_{13}^+ &  p_{13}^-\\
p_{21}^+ & p_{21}^- &  p_{22}^+ &  p_{22}^- & p_{23}^+ &  p_{23}^-\\
p_{31}^+ & p_{31}^- &  p_{32}^+ &  p_{32}^- & p_{33}^+ &  p_{33}^-\\
d_{11}^+ & d_{11}^- &  d_{12}^+ &  d_{12}^- & d_{13}^+ &  d_{13}^-\\
d_{21}^+ & d_{21}^- &  d_{22}^+ &  d_{22}^- & d_{23}^+ &  d_{23}^-\\
d_{31}^+ & d_{31}^- &  d_{32}^+ &  d_{32}^- & d_{33}^+ &  d_{33}^-
\end{bmatrix}
\quad
\operatorname{diag} \mathrm{e}^{\mathrm{i}(\pm k\alpha_i h)}
\quad
\begin{bmatrix}
A_{1}^{+}\\
A_{1}^{-}\\
A_{2}^{+}\\
A_{2}^{-}\\
A_{3}^{+}\\
A_{3}^{-}
\end{bmatrix}
\end{equation}
with $\sigma_{i3}^* = \frac{\sigma_{i3}}{\mathrm{i}k}$. For a single layer $n$ this equation is written in compact form as
\begin{equation}
\mathbf{U}_{n} = \mathbf{G}_{n}\mathbf{H}_{n}\mathbf{A}_{n}.
\label{eqn:Gesamt1}
\end{equation}
With this framework at hand, the continuity condition for the stresses and displacements at the layer interfaces
\begin{equation}
\mathbf{U}_{n}^{+} - \mathbf{U}_{n+1}^{-} = 0,
\label{eqn:Konti}
\end{equation}
where ${U}_{n}^{+}$ gives the values at the upper surface of the layer $n$ and ${U}_{n+1}^{-}$ at the lower surface of the layer $n+1$, respectively, provides the dispersion relation for GUW in layered structures like FML
\begin{equation}
\begin{vmatrix}
[\mathbf{G}_{(N)}\mathbf{H}_{(N)}^{+}]_{4-6}  & & & 0 \\
\mathbf{G}_{(N)}\mathbf{H}_{(N)}^{-}& -\mathbf{G}_{(N-1)}\mathbf{H}_{(N-1)}^{+} & & \\
&  \ddots & \ddots &  \\
  & & \mathbf{G}_{(2)}\mathbf{H}_{(2)}^{-}& -\mathbf{G}_{(1)}\mathbf{H}_{(1)}^{+}  \\
0  & & & \mathbf{G}_{(1)}\mathbf{H}_{(1)}^{+}]_{4-6}
\end{vmatrix}
=0.
\label{eqn:Transfer6}
\end{equation}  
The subscript $4-6$ refers to lines 4 to 6 of Eq.~(\ref{eqn:Gesamt1}), which give the stress-free boundary conditions at the upper and lower edge of the layered structure. 
\par 
It is important to note that there is no closed solution for this set of equations. Thus, the computation of dispersion diagrams is based on an iterative procedure utilizing the bisection method. 
Even though the results are determined numerically, this solution will be referred to as an analytical solution in the course of this work. The aim is to clearly indicate the difference to a pure numerical solution on the basis of the FEM.

\section{Experimental setup}
In this section, the experimental setup for the determination of the dispersion diagrams is presented. This includes the specifications and manufacturing of the specimens, the explanation of two different experimental setups (ES1 and ES2) as well as the data acquisition and the postprocessing procedure.

\subsection{Materials and specimen manufacturing process}
\label{sec:materials_and_specimen}
The specimens in this work are made of alternating CFRP prepreg (Hexcel Hexply 8552-AS4) layers and thin steel foils (1.4310). Table~\ref{tab:material_properties_CFRP} presents the material properties of the CFRP prepreg from different literature sources. The properties from NCAMP~\cite{NCAMP.2011} are best documented in terms of experimental procedures and test results. It should be noted, that the tests were performed with a prepreg material with a higher fiber area weight and a slightly different fiber volume content compared to the data sheet~\cite{HexcelCorporation.2016} of the prepreg at hand. In Garstka~\cite{Garstka.2005} the value of $G_{23}$ is given equal to $G_{12}$ and $G_{13}$ which is not in accordance to the rules for transversal isotropic materials. Therefore, the value of $G_{23}$ is questionable. Hörberg~\cite{Horberg.2019} gives the same values but with an updated $G_{23}$. Here the material is not specified by the authors but is assumed to be Hexcel 8552-AS4 due to the agreement of the other values. Johnston~\cite{Johnston.1997} gives the most comprehensive set of parameters with a fiber volume content similar to the data sheet of the material in this work. However, the tensile modulus in fiber direction shows around \SI{10}{\percent} difference to the values in other literature sources. The material properties for the stainless steel alloy 1.4310 are given in Table~\ref{tab:material_properties_steel}. Own tensile tests of the steel foil show slightly higher values for the Young's modulus compared to the values provided by the DIN norm~\cite{DIN.2003}.
\par
\begin{figure}
\begin{floatrow}
\ffigbox[0.99\linewidth]
{\includegraphics[width=\linewidth]{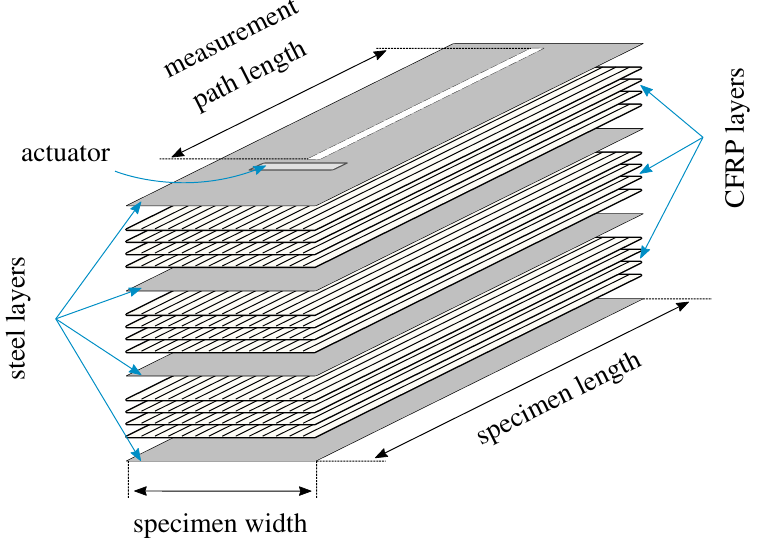}}
{\caption{Schematic representation of the specimen architecture.}
\label{fig:specimen_definition}}
\ffigbox[0.99\linewidth]
{\includegraphics[width=\linewidth]{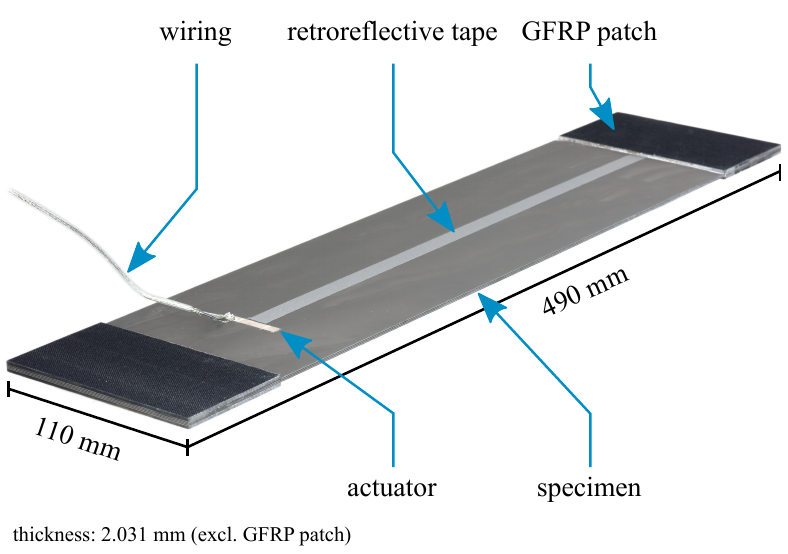}}
{\caption{Photo of the used strip specimen.}
\label{fig:photo_strip_specimen}}
\end{floatrow}
\end{figure}

The specimens used in this work consist of 4 metal layers and 12 CFRP layers that are placed in a symmetric layup with two metal layers at the top and bottom of the specimen. In laminate notation, the layup can be described as $[St/0_4/St/0_2]_S$. With the single-ply thicknesses from Table~\ref{tab:material_properties_CFRP} and~\ref{tab:material_properties_steel}, this layup results in a nominal laminate thickness of \SI{2.04}{\milli \meter} and a metal volume fraction (MVF) of \SI{24}{\percent}. The layup and specimen architecture are schematically illustrated in Figure~\ref{fig:specimen_definition}.

\begin{table}
    \small\sf\centering
	\caption{Comparison of material properties for the used prepreg material Hexcel Hexply 8552-AS4 from literature.}
	\label{tab:material_properties_CFRP}
	\begin{tabular}{lcccccc}
	    \toprule
		Value & Unit & NCAMP~\cite{NCAMP.2011} & Garstka~\cite{Garstka.2005} & Hörberg~\cite{Horberg.2019}& Johnston~\cite{Johnston.1997}& Data sheet~\cite{HexcelCorporation.2016} \\
		\midrule
		$E_{1}$ & \si{\giga \pascal} & 132 & 135 &135& 122 &141 \\ 
		$E_{2} =E_{3}$ & \si{\giga \pascal} & 9.2 & 9.5 &9.5& 9.9 &10 \\ 
		$G_{12} =G_{13}$ & \si{\giga \pascal} & 4.8 & 4.9 &4.9& 5.2 &- \\ 
		$G_{23}$ & \si{\giga \pascal} & - & \textit{(4.9)}$^a$ &3.3& 3.4 &-\\
		$\nu_{12} =\nu_{13}$ & -  & 0.3 & 0.3 &0.3& 0.27 &-\\ 
		$\nu_{23} =\frac{E_2}{2G_{23}}-1$ & -  & - & 0.45 &0.45& 0.47  \\ \midrule
		f. vol. content & \% & 60.38 & 57.4& -& 57.3 & 57.42    \\
        f. area weight & \si{\gram \per \meter \squared} & 190 & - & - & - & 134  \\
		$\rho$ & \si{\gram \per \cubic \centi \meter} & 1.57-1.60 & - & - & -&1.58  \\ 
		$t_{ply}$ (cured) & \si{\milli\meter} & 0.19 & - & 0.13 & -& 0.13 \\ \bottomrule
		\multicolumn{7}{l}{\footnotesize{$^\textrm{a}$ not in accordance with rules of transversal isotropy}} \\
	\end{tabular}
\end{table}

\begin{table}
    \small\sf\centering
    \caption{Material properties for the austenitic steel alloy 1.4310 (X10CrNi18-8).}
	\label{tab:material_properties_steel}
	\begin{tabular}{lccc}
		\toprule
		Value & Unit &  DIN EN 10151 \cite{DIN.2003}& own pretests\\
		\midrule
		$E$ & \si{\giga \pascal} & 179 & 191\\ 
		$G$ & \si{\giga \pascal} &  68.8$^\textrm{b}$ & 73.5$^\textrm{b}$ \\
		$\nu$ & - & 0.3 & -\\
		$t_{ply}$& \si{\milli\meter} & 0.12 & 0.12 \\ \bottomrule
		\multicolumn{4}{l}{\footnotesize{$^\textrm{b}$ calculated for linear elastic isotropic materials: $G=\frac{E}{2(1+\nu)}$}} \\
	\end{tabular}
\end{table}

The process for the specimen production is a standard process for FMLs with fiber layers from prepreg systems. To assure a high interlaminar bonding strength between the different FML layers, the steel is mechanically pretreated with a vacuum suction blasting process. This process is beneficial for thin materials that would be damaged by classic sandblasting~\cite{Stefaniak.2012b}. After the mechanical treatment, the surface is chemically cleaned with heptane. Subsequently, an aqueous sol-gel solution (3M Surface Pre-Treatment AC-130-2~\cite{3M.2015}) is applied to all metal surfaces. The sol-gel layers increase the adhesive strength between CFRP and metal and lead to high interlaminar shear strength of the FML~\cite{Stefaniak.2017}. The metal foil is dried for one hour after the sol-gel application, according to the data sheet, and laminated immediately together with the CFRP prepreg layers~\cite{Blohowiak.1999, Stefaniak.2017}.
\par
To cure the FML, the standard manufacturer-recommended autoclave process as found in the CFRP datasheet~\cite{HexcelCorporation.2016} is used. The process consists of a dwell stage at \SI{110}{\degreeCelsius} and a final cure temperature of \SI{180}{\degreeCelsius}.
\par

In this work, two specimens are investigated, which mainly differ in their dimensions. One is strip-shaped, the other has the shape of a square plate. The dimensions of the specimens as well as the length of the resulting measuring paths are given in Table~\ref{tab:specimen_geometries}.

The strip specimen with actuator and retroreflective tape is depicted in Figure~\ref{fig:photo_strip_specimen}. It is equipped with patches of GFRP at both ends outside of the measuring path for future tests in a tensile testing machine. In order to reduce the influence of occurring edge reflections, the strip edges along both sides are prepared with an energy dissipating butyl rubber tape.
Since no GFRP patches are used for the plate specimen, the available measurement path length is larger compared to the strip specimen. This is exploited in the measurements in ES2 to increase the quality of the measurements. In ES1, however, the measuring length for the plate specimen is reduced to the maximum path length of the strip specimen. This provides better comparability between the measurements of the two specimen types.

A rectangular piezoceramic actuator (ceramic type: PIC 255) with a length of \SI{30}{\milli \meter}, a width of \SI{5}{\milli \meter} and a ceramic thickness of \SI{0.2}{\milli \meter} is used. This shape results in a more straight wavefront along the measuring path in comparison to a circular shaped actuators. The actuator contains a wrap-around electrode for single-sided access of the wiring. It is bonded to the specimen with Loctite EA9466~\cite{Henkel.2019}, which was cured under a vacuum bag to ensure a homogeneous thickness of the adhesive layer.

\begin{table}[htbp]
	\small\sf\centering
	\caption{Geometry and measurement path length of the two investigated specimens.}
	\label{tab:specimen_geometries}
	\begin{tabular}{lcc}
		\toprule
		 & Strip & Plate \\
		 & [\si{\milli \meter}] & [\si{\milli \meter}] \\
		\midrule
		Specimen length             & 490                           & 500  \\
		Specimen width              & 110                           & 500 \\
		Measurement path length ES1 & 320                           & 320 \\
		Measurement path length ES2 & 320                           & 450 \\
		\bottomrule
	\end{tabular}
\end{table}

The laminate thickness is a crucial parameter for the subsequent calculation of the GUW propagation. Therefore, the specimen thickness and its distribution over the surface of the specimens is experimentally determined. For each specimen, the thickness is measured with a dial gauge after manufacturing at 10 evenly distributed points over its surface. Table~\ref{tab:specimen_thickness} shows that only minor deviations to the nominal thickness of \SI{2.04}{\milli \meter} can be found.

\begin{table}[htbp]
	\small\sf\centering
    \caption{Mean values resulting from thickness evaluations at 10 evenly distributed points over the surface for the plate and strip specimen.}
	\label{tab:specimen_thickness}
	\begin{tabular}{lcc}
		\toprule
		 & Strip & Plate \\ 
		 & [\si{\milli \meter}] & [\si{\milli \meter}] \\
		\midrule
		Mean thickness & 2.031 & 2.046 \\
		Standard deviation & 0.003 & 0.014 \\
        \bottomrule
	\end{tabular}
\end{table}

Additionally, the strip specimen is evaluated by means of an industrial 3d scanning head (GOM ATOS) to further evaluate the homogeneity of the manufactured laminate thickness.

\begin{figure}
\begin{floatrow}
\ffigbox[0.99\linewidth]
{\includegraphics[trim={1.6cm 13.3cm 1.6cm 4.6cm},clip, width=\linewidth]{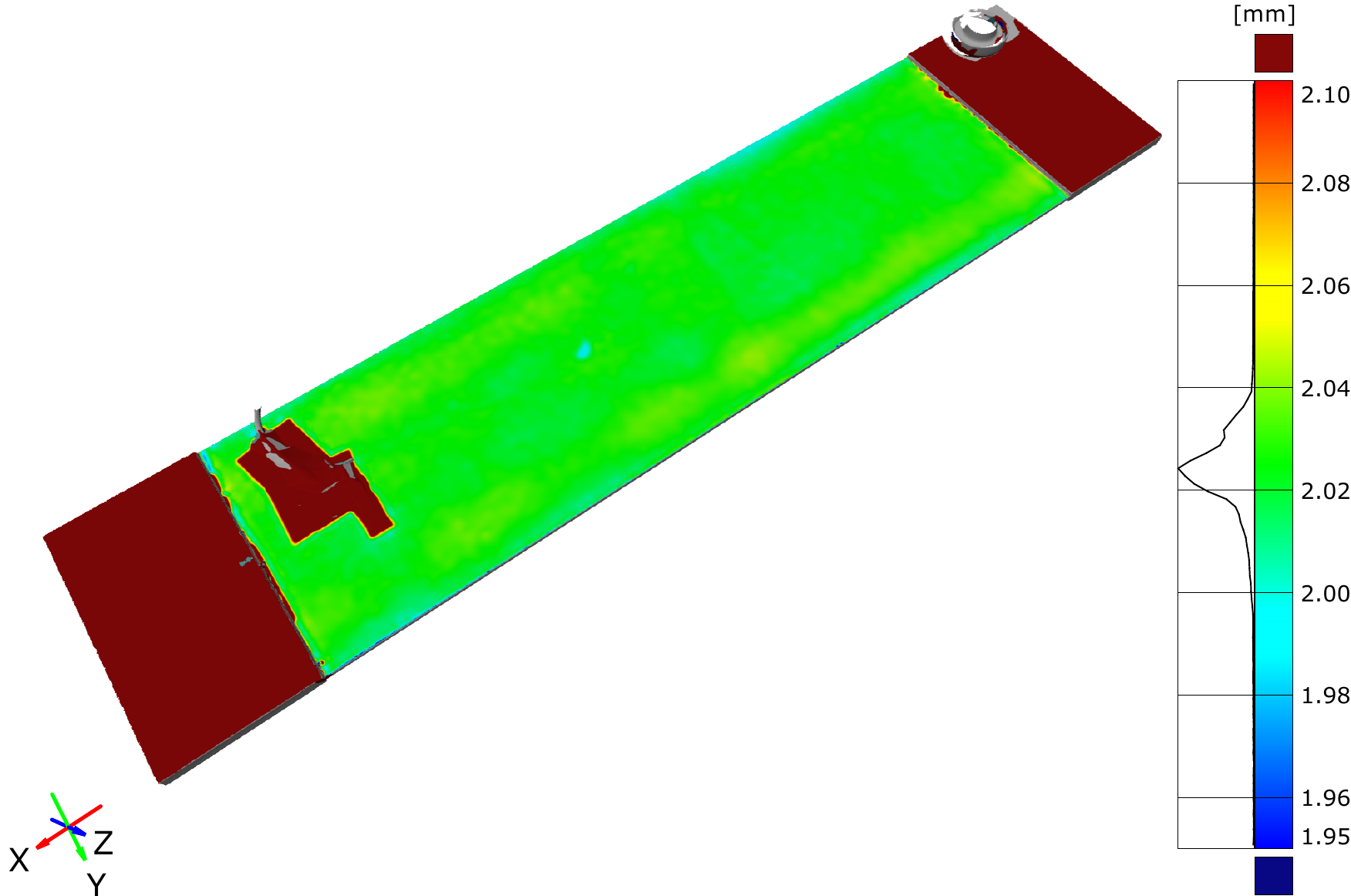}}
{\caption{Evaluation of the thickness homogeneity within the strip specimen by 3d scanning.}
\label{fig:GOM_ATOS_strip_thickness}}
\end{floatrow}
\end{figure}

The 3d representation in Figure~\ref{fig:GOM_ATOS_strip_thickness} shows a very evenly distributed thickness over the entire strip with the same range as the results from the dial gauge measurements (cf. Table~\ref{tab:specimen_thickness}). Both measurement methods include their own measuring error. However, both results are very close to the nominal specimen thickness of \SI{2.04}{\milli \meter} which is therefore used throughout this work.

\subsection{Equipment and setup}
\label{sec:exp_setup}
The experimental setup aims at measuring the wave velocity at the presented specimen's surface along the wave propagation direction using a full-field LSV, cf.~\cite{Lammering.2018,Neumann.2013,Barth.ExpM.2022, Thierry.2020}. 
\par
\begin{figure}
\begin{floatrow}
\ffigbox[0.99\linewidth]
{\includegraphics[width=\linewidth]{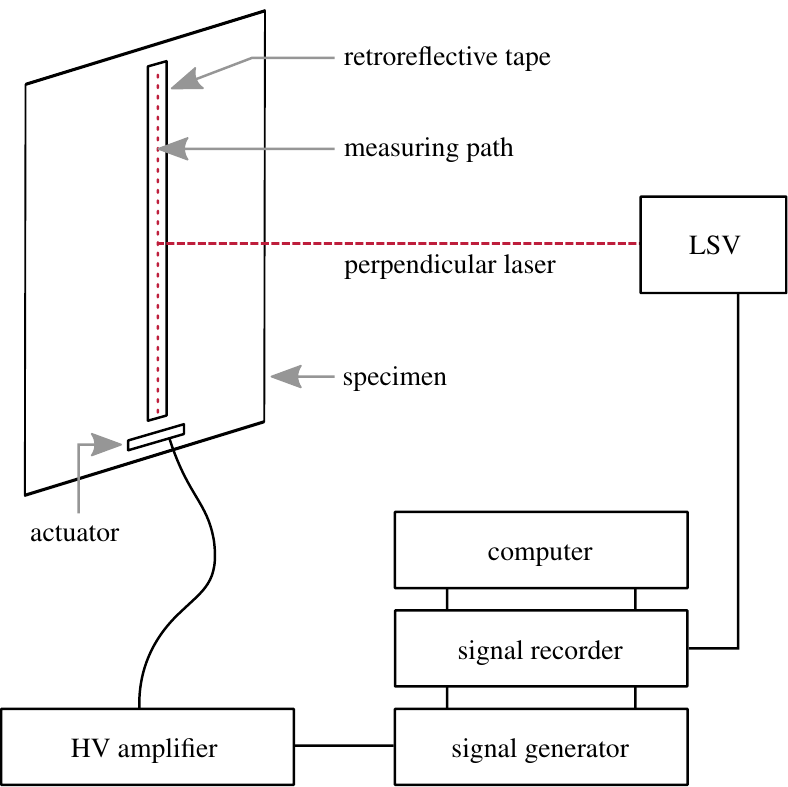}}
{\caption{Schematic representation of the experimental setup.}
\label{fig:exp_setup_sketch}}
\end{floatrow}
\end{figure}

Figure~\ref{fig:exp_setup_sketch} shows a schematic representation of the setup. A computer in combination with a signal generator provides the excitation signal. The excitation signal is amplified using a high-voltage (HV) amplifier to drive the actuator that is bonded to the FML specimen. The specimens are mounted vertically in such a way that no clamping on the vertical edges takes place and that an upright position is ensured at all times. Retroreflective tape is applied along the measurement path to increase the received signal amplitude by the LSV. Furthermore, a perpendicular alignment of the laser beam to the center of the measurement path is ensured. Taking into account the laser specifications, the horizontal distance between the LSV and the specimen is selected in such a way that both, the highest possible laser intensity can be measured and the spatial sampling along the measuring path is as high as possible. The velocities at the specimen's surface are acquired using a signal recorder and post-processed with the computer. Due to the positioning and orientation of the LSV, almost exclusively the out-of-plane component of the wave velocities is measured. 
\par
\begin{figure}
\begin{floatrow}
\ffigbox[0.99\linewidth]
{\includegraphics[width=\linewidth]{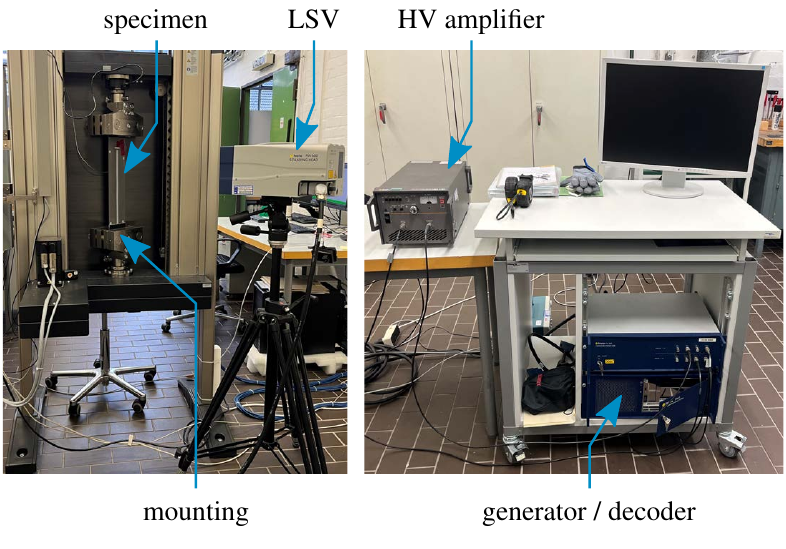}}
{\caption{Experimental setup 1 (ES1)}
\label{fig:exp_setup_hh}}
\ffigbox[0.99\linewidth]
{\includegraphics[width=\linewidth]{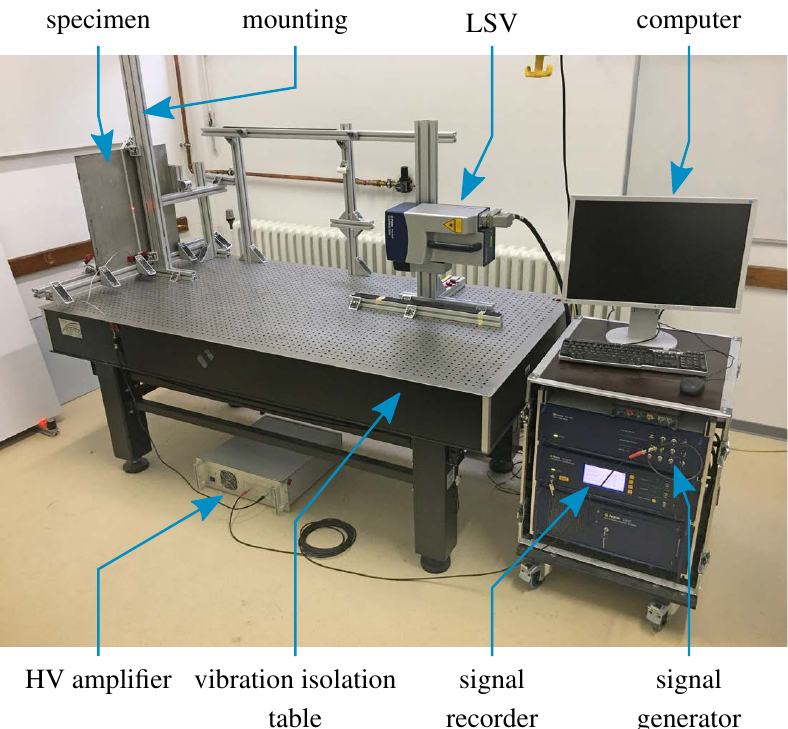}}
{\caption{Experimental setup 2 (ES2)}
\label{fig:exp_setup_bs}}
\end{floatrow}
\end{figure}

In the context of data transfer in research cooperations such as the authors' research unit which is located at different institutes, and general applicability of the method under investigation, the question arises, whether it works independently of the setup. The method is assumed to perform with different setup specifications as long as the temporal and spatial resolutions as well as the measuring durations and path lengths correspond to the signal processing requirements.
Therefore, the measurements are carried out using two setups with the same architecture but different hardware. The two experimental setups are depicted in Figure~\ref{fig:exp_setup_hh} (ES1) and Figure~\ref{fig:exp_setup_bs} (ES2), respectively. A description of the hardware of the different setups is presented in Table~\ref{tab:hardware_comparison}. The decisive differences between the two measurement setups for this application are the sampling rate of the signal generators and the accuracy of the angular resolution of the LSVs used. The signal generator in ES2 has a maximum sampling rate of \SI{1}{\mega \hertz} which, according to Shannon~\cite{Shannon.1949}, results in a maximum excitation frequency of \SI{0.5}{\mega \hertz}, whereas for ES1 a much higher excitation frequency of \SI{62.5}{\mega \hertz} is theoretically possible. In the context of this work, a maximum excitation frequency of \SI{1}{\mega \hertz} is used for ES1. Further, the accuracy of the angular resolution in ES2 affects the accuracy of signals with high wavenumbers. The resulting influence is discussed in Section~\ref{sec:data_sorting}.

\begin{table}
\setlength{\tabcolsep}{4pt}
    \small\sf\centering
	\caption{Comparison of hardware specifications used in ES1 and ES2.}
	\label{tab:hardware_comparison}
	\begin{tabular}{lccc}
		\toprule
		Device and parameter & ES1 & ES2 & Unit\\
		\midrule
		\textbf{Vibrometer} &  &  & \\
		\hspace{.1cm} Manufacturer & Polytec & Polytec & -\\
		\hspace{.1cm} Model & PSV-500 & PSV-400 & -\\
		\hspace{.1cm} angular resolution & $<$ 0.001 & 0.002 & \si{\degree} \\
		\hspace{.1cm} Used working distance & $\approx535$ & $\approx1330$ & \si{\milli\metre} \\
		\midrule
		\textbf{Signal generator} &  & & \\
		\hspace{.1cm} Manufacturer & Polytec & Polytec & -\\
		\hspace{.1cm} Bandwidth & 30 & 0.5 & \si{\mega \hertz}\\
		\hspace{.1cm} No. of sampling points & \SI{1e6}{} & \SI{1.3e5}{} & -\\
		\hspace{.1cm} Max. sampling rate & 125 & 1 & \si{\mega \hertz} \\
		\hspace{.1cm} Used sampling rate & 10 & 1 & \si{\mega \hertz} \\
        \midrule
		\textbf{Signal recorder} &  &  &  \\
		\hspace{.1cm} Manufacturer & Polytec & Polytec & -\\
		\hspace{.1cm} Sampling frequency & 3.125 & 2.56 & \si{\mega\hertz}\\
		\hspace{.1cm} Used resolution & 16 & 12 & bit\\
		\hspace{.1cm} Used amplitude range & 100 & 20 & \si{\milli\metre/\second/\volt}\\
		\bottomrule
	\end{tabular}
\end{table}
Within this work, measurement labels are used to distinguish the results between different setups and specimens, cf. Table~\ref{tab:measurement_labels}. The label distinguishes between the two experimental setups ES1 and ES2, the specimen types plate (P) and strip (S) as well as the measurement runs MS1 and MS2 if applicable.
\begin{table}[htbp]
	\small\sf\centering
	\caption{Description of measurement labels.}
	\label{tab:measurement_labels}
	\begin{tabular}{lccc}
		\toprule
		Label & Setup & Type & Run\\ 
		\midrule
		ES1.P & 1 & plate & 2\\
		ES1.S.MS1 & 1 & strip & 1\\
		ES1.S.MS2 & 1 & strip & 2\\
		ES2.S & 2 & strip & 1\\
		ES2.P.MS1 & 2 & plate & 1\\
		ES2.P.MS2 & 2 & plate & 2\\
        \bottomrule
	\end{tabular}
\end{table}

\subsection{Data acquisition and postprocessing}
\label{sec:exp_post}
The experimental data acquisition and postprocessing used for the determination of phase velocities of the GUWs is based on Barth et al.~\cite{Barth.ExpM.2022}. It is characterized by very high accuracy, reproducibility, and a possibility for automation, which allows the determination of the dispersion relations over a wide frequency range. The method is based on the use of a non-uniform 2d-DFT to evaluate the measurement data for GUW. The successful use of a 2d-DFT for this type of applications has been demonstrated before by~\cite{Cawley.1991,Hora.2012,Su.2009}.
\par
\begin{figure}
\begin{floatrow}
\ffigbox[0.99\linewidth]
{\includegraphics[width=\linewidth]{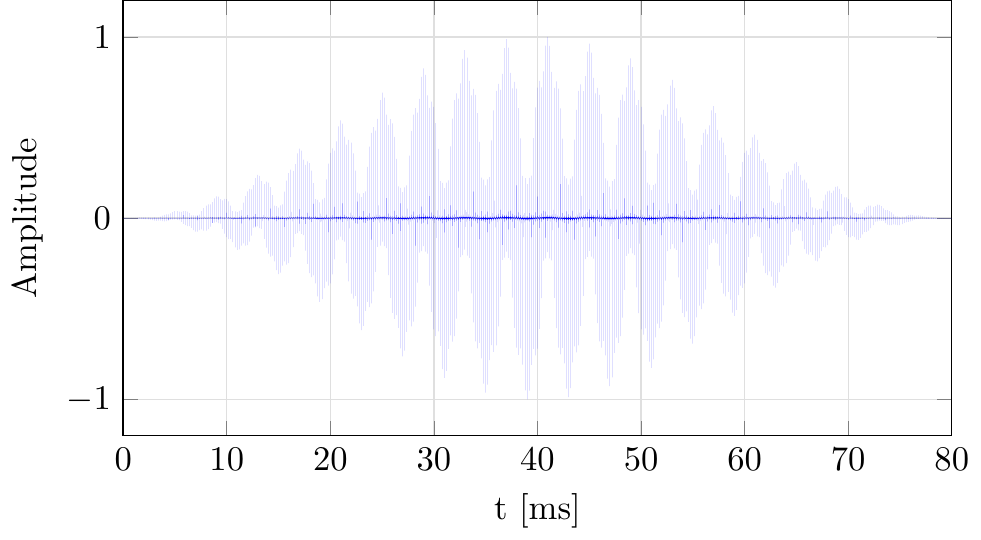}}
{\caption{Exemplary representation of a normalized multifrequency excitation signal in the time domain.}
\label{fig:anregung_time}}
\ffigbox[0.99\linewidth]
{\includegraphics[width=\linewidth]{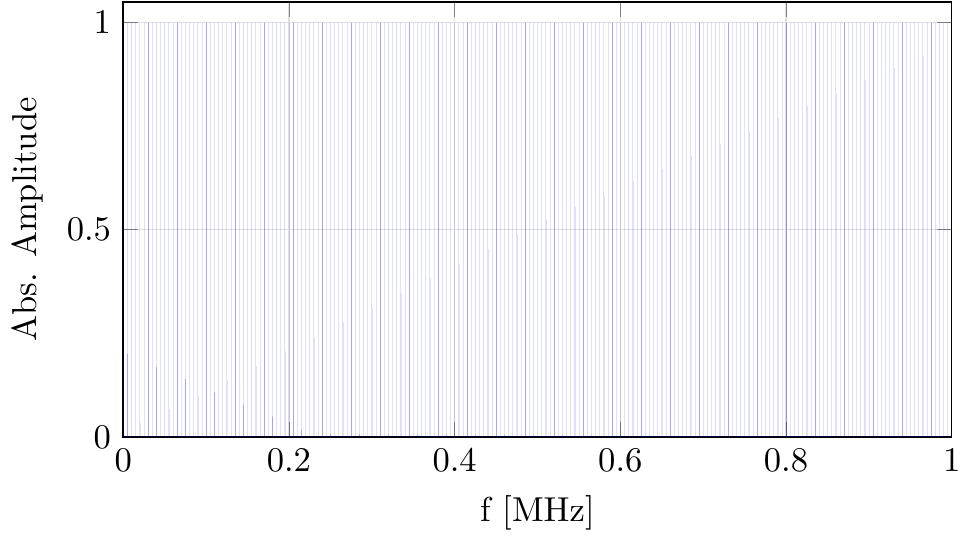}}
{\caption{Normalized multifrequency excitation signal from Figure~\ref{fig:anregung_time} in the frequency domain.}
\label{fig:anregung_freq}}
\end{floatrow}
\end{figure}

The evaluation of the measurement data is done in the frequency-wavenumber domain instead of the often used evaluation in the time-space domain. Since at least two modes occur in case of monofrequency excitation~\cite{Lamb.1917,graff.2012,Achenbach.1973,Prosser.1999,Staszewski.2004}, the frequency-wavenumber domain is well suited for the analysis of the dispersive and multimodal behavior of such waves. This also allows the use of multifrequency excitation signals to automate the measurement, which is hardly possible when the data is evaluated in the time-space domain. An exemplary excitation signal in the time domain is shown in Figure~\ref{fig:anregung_time} and the transformed signal in the frequency domain in Figure~\ref{fig:anregung_freq}. It consists of a Hanning windowed~\cite{blackman.1959} superposition of sinusoidal oscillations and has a temporal length of $T=$ \SI{80}{\milli \second} or \SI{125}{\milli \second} for the two experimental setups, respectively. The duration can be considered as untypically long for GUW applications, where mostly short bursts or impulse excitations are used. The signal is chosen to ensure a very accurate resolution in the frequency domain.
In the first test run, frequencies from \SI{0.25}{\kilo\hertz} to \SI{995.25}{\kilo\hertz} are excited with a step size of \SI{5}{\kilo\hertz}. By repeating the measurement 20 times with excitation signals shifted by \SI{0.25}{\kilo\hertz}, a range from $f_{min}=$ \SI{0.25}{\kilo\hertz} to $f_{max}=$ \SI{1}{\mega\hertz} with a step size of $\Delta f=$ \SI{0.25}{\kilo\hertz} is covered.
The excitation signal, described and shown in Figures~\ref{fig:anregung_time} and~\ref{fig:anregung_freq}, is applied in ES1 and is limited to a smaller frequency range and a wider spacing for ES2 due to hardware limitations (cf. Table~\ref{tab:hardware_comparison}). Table~\ref{tab:prop_exc} shows the properties of the investigated frequency range for both experimental setups.\\

\begin{table}[htbp]
	\small\sf\centering
	\caption{Properties of investigated frequency ranges for ES1 and ES2.}
	\label{tab:prop_exc}
	\begin{tabular}{lcccc}
		\toprule
		Setup & $f_{min}$ & $f_{max}$ & $\Delta f$ &T\\ 
		& [\si{\kilo \hertz}]  & [\si{\mega \hertz}]&[\si{\kilo \hertz}]&[\si{\milli \second}]\\ 
		\midrule
		ES1 & 0.25 & 1 & 0.25 & 80\\
		ES2 & 1 & 0.5 & 1 & 125\\
        \bottomrule
	\end{tabular}
\end{table}

As described in Section~\ref{sec:exp_setup}, the required  discrete-time velocity data $\textbf{c}(t,x)$ is generated using a LSV measurement on the specimen's surface at distributed points along a path in the propagation direction $x$ of the waves. The resulting data are processed using a non-uniform 2d-DFT \cite{bracewell.fourier.1986} 
\begin{equation}
\mathbf F \left( f,\tilde{\nu} \right) = \sum_{x=0}^n \sum_{t=0}^m \left( \textbf{c}(t,x)  e^{ i  2 \pi f t}\right) e^{- i  2 \pi  \tilde{\nu} x} \quad ,
\label{eqn:dft}
\end{equation}
which converts the data into the frequency-wavenumber domain. Here, $f$ signifies the frequency and $\tilde{\nu}$ the wavenumber. Within the amplitude matrix $\mathbf F$ of Equation~(\ref{eqn:dft}), maxima are detected at given frequency points using a peak-search algorithm. Each resulting frequency-wavenumber pair represents a detected point in the Lamb wave dispersion relation.
\par
In contrast to a direct evaluation of the time signals by e.g. time-of-flight measurements, the advantage of this procedure is a much easier separation of the modes in the frequency-wavenumber domain. In addition, it is possible to determine real phase velocities instead of the often measured group velocities, which is not possible with many other measurement methods without great effort or inaccuracies.

\subsection{Identification of outliers}
\label{sec:data_sorting}
\begin{figure}
\begin{floatrow}
\ffigbox[0.99\linewidth]
{\includegraphics[width=\linewidth]{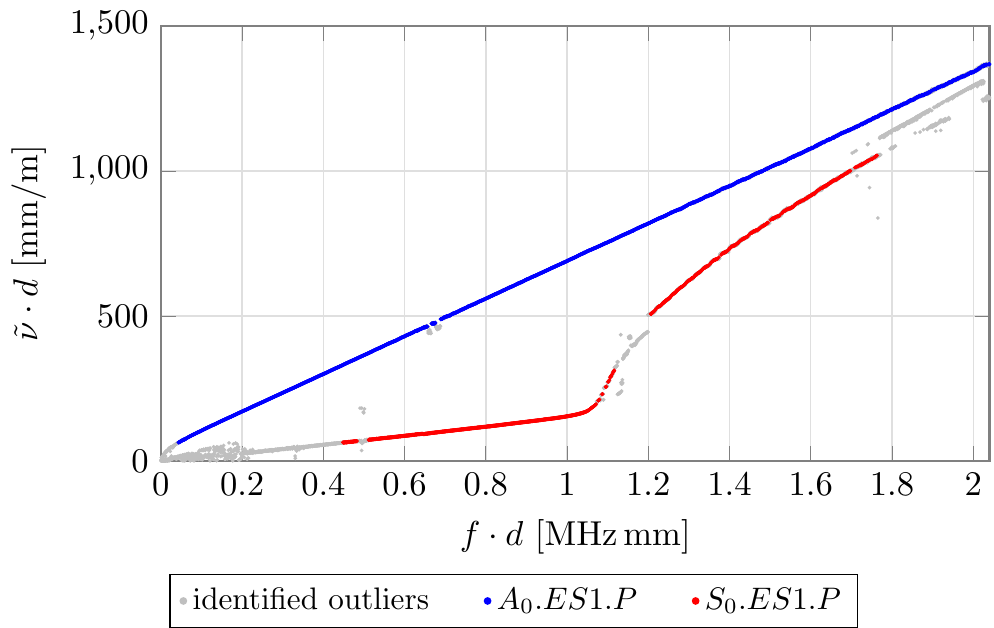}}
{\caption{Exemplary outlier identification for dispersion diagrams in frequency wavenumber representation.}
\label{fig:sort_data}}
\ffigbox[0.99\linewidth]
{\includegraphics[width=\linewidth]{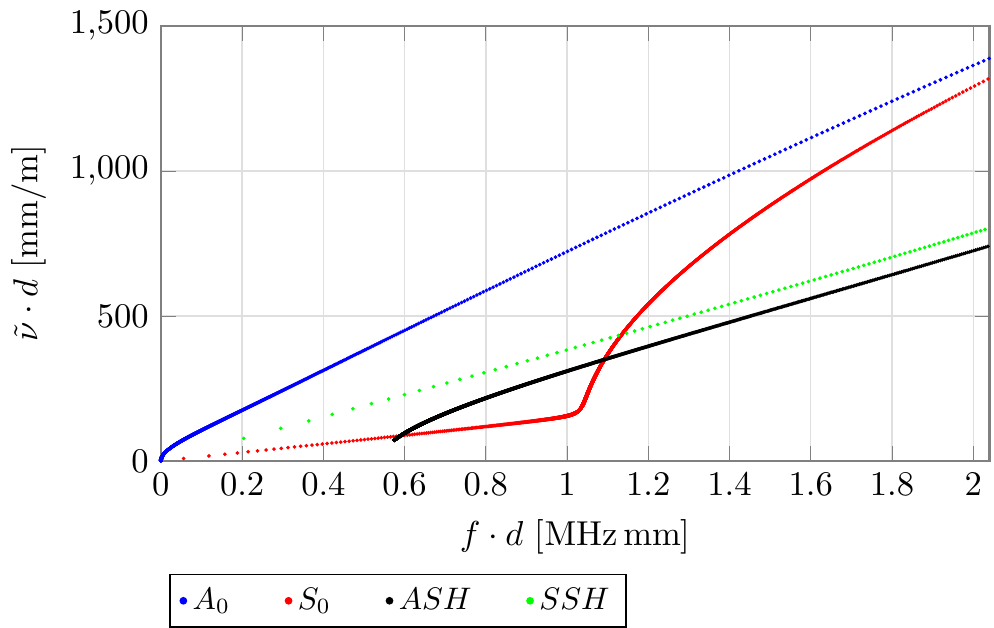}}
{\caption{Dispersion diagram from analytical framework including antisymmetric (ASH) and symmetric (SSH) shear horizontal waves. Material parameters from~\cite{Johnston.1997}.}
\label{fig:sh_analytic}}
\end{floatrow}
\end{figure}

As expected, the measurements performed partially contain erroneous data due to external disturbances and setup- or specimen-specific characteristics. To make the data more comparable, the outliers are detected as exemplarily shown in Figure~\ref{fig:sort_data}. 
The identification of outliers is carried out in the frequency-wavenumber domain. This is done because of the large linear course of the data in this domain. In contrast, the course is mostly non-linear in the phase velocity domain.
Two main criteria are used to identify and eliminate the outliers: 
\par
\begin{itemize}
\item All measurement data below and above a specified wavenumber are sorted out. The minimum wavenumber depends on the length of the measuring path of the respective specimen as described in Barth et al.~\cite{Barth.ExpM.2022}. It is required that the measurement path length must have at least ten times the length of the maximum wavelength, i.e. minimum wavenumber. The maximum wavenumber depends on the distance between successive measuring points. It is assumed that the spatial sampling frequency must be at least twice as high as the highest wavenumber. The resulting minimum and maximum frequencies are shown in Table~\ref{tab:freq_ranges}.
\item Measurement data is sorted out by generating a fit of the measurement data and determining outliers employing a specified maximum error residual. In the wavenumber domain, a linear fit function can be used for most measurement ranges and thus the desired outliers can be detected through a single step. In the frequency ranges in which a linear function is not sufficient, a spline fit is used, which approximates the function step by step and sorts out outliers. For this, the number of measuring points in a certain area is used as weighting for the used spline functions.
\end{itemize}
Additional reasons for sorting out data because of reappearing measurement deviations are listed below:
\begin{itemize}
\item Due to a more inaccurate angular resolution in the LSV at ES2 compared to ES1, data with high wavenumbers had to be removed from consideration since a high scattering of the results occurred in this range. The more prominent occurrence of this effect in the measurement of the strip specimen (ES2.S) compared to the ones in the plate specimen (ES2.P.MS1 and ES2.P.MS2) can be attributed to the shorter measuring path, see Table~\ref{tab:specimen_geometries}, since a longer measuring path can compensate for this effect to some degree.
\item In the data sorting of the $S_0$-mode between \SI{1.1}{\mega \hertz \milli \meter} and \SI{1.2}{\mega \hertz \milli \meter} a large number of outliers are particularly striking, since they are repeated in all data sets. To explain this effect, the numerical solution of the analytical framework of the propagating waves must be considered. That this solution is applicable here will be shown in the course of the paper in Section~\ref{sec:comparison_analytic_experiment}. The deviations can be explained by the occurring shear horizontal wave modes ($SH$-modes) in this frequency range. Figure~\ref{fig:sh_analytic} shows a representation based on the numerical solution of the analytical framework. It can be seen that two $SH$-modes are superposing the $S_0$-mode in this range. Here, $ASH$ and $SSH$ stand for an antisymmetric and symmetric shear horizontal wave. This leads to interferences between the wavemodes and causes a disturbance in the evaluation of the $S_0$-mode. It should be noted that the frequency range of the perturbation differs slightly between the experimental data in Figure~\ref{fig:sort_data} and the analytical data in Figure~\ref{fig:sh_analytic}. A detailed discussion on the comparison of the experimental data and the numerical solutions of the analytical framework is given in Section~\ref{sec:comparison_analytic_experiment}.

\item At frequencies above \SI{1.8}{\mega \hertz \milli \meter} the wavenumbers of the two fundamental GUW modes converge more and more. This results in a superposition of the amplitude maxima in the frequency-wavenumber matrix $\mathbf F \left( f,\tilde{\nu} \right)$ during the detection and finally in an erroneous shift in the detection of both modes. However, the shift is especially visible in the $S_0$-mode, since the out-of-plane velocity has considerably smaller amplitudes than for the $A_0$-mode.
\end{itemize}
\par

\begin{table}[htbp]
	\small\sf\centering
	\caption{Frequency-thickness ranges for different experimental setups and measurements.}
	\label{tab:freq_ranges}
	\begin{tabular}{lcccc}
		\toprule
		Setup & Mode & $(f \cdot d)_{min}$ & $(f \cdot d)_{max}$ & $\Delta (f \cdot d)$\\ 
		&  & [\si{\kilo \hertz \milli \meter}]&[\si{\mega \hertz \milli \meter}]&[\si{\kilo \hertz \milli \meter}]\\ 
		\midrule
		ES1.S & $A_0$ & 43 & 2.04 & 0.51 \\
		ES1.S & $S_0$ & 443 & 2.04 & 0.51 \\
		ES1.P & $A_0$ & 43 & 2.04 & 0.51 \\
		ES1.P & $S_0$ & 443 & 2.04 & 0.51 \\
		\midrule
		ES2.S & $A_0$ & 43 & 0.68 & 2.04 \\
		ES2.S & $S_0$ & 443 & 1.02 & 2.04 \\
		ES2.P & $A_0$ & 26 & 0.96 & 2.04 \\
		ES2.P & $S_0$ & 324 & 1.02 & 2.04 \\
        \bottomrule
	\end{tabular}
\end{table}
\par

\section{Results}
\label{sec:results}
In the following, the results of the measurements are compared with each other and with the numerical solution of the analytical framework based on various criteria. 
The results are structured as follows. Sections~\ref{sec:reproducibility} and \ref{sec:reproducibility_ES} address the reproducibility and comparability of the measurements. Thereby, the former considers the reproducibility with the same experimental setup while the latter compares the comparability between different experimental setups. The influence of the specimen width on the measurement results is described in Section~\ref{sec:comparison_specimen_size}. A conclusion of the considerations is offered in Section~\ref{sec:comparison_analytic_experiment} where a comparison of the measured data to the results of the analytical framework is given. 
\par
It should be noted that the measurement data, as described in Section~\ref{sec:exp_post} are available in the frequency-wavenumber domain. However, to ensure better comparability with other publications, all subsequent comparisons are made using the phase velocity $c_p = f / \tilde{\nu}$. 

\subsection{Validating the measurement method with the same experimental setup}
\label{sec:reproducibility}
In this subsection, the reproducibility of the measurement results is investigated separately for both experimental setups. This is done to check for the method's reliability when generating GUW dispersion diagrams in FML. For each setup, two independent measurements are performed. The specimen is reinstalled and the LSV is moved between the two runs. Moreover, the method is applied to a strip and a plate specimen, respectively, to show reproducibility for different specimen geometries.

\subsubsection{Comparison of strip specimen measurements with ES1}
\begin{figure}
\begin{floatrow}
\ffigbox[0.99\linewidth]
{\includegraphics[width=\linewidth]{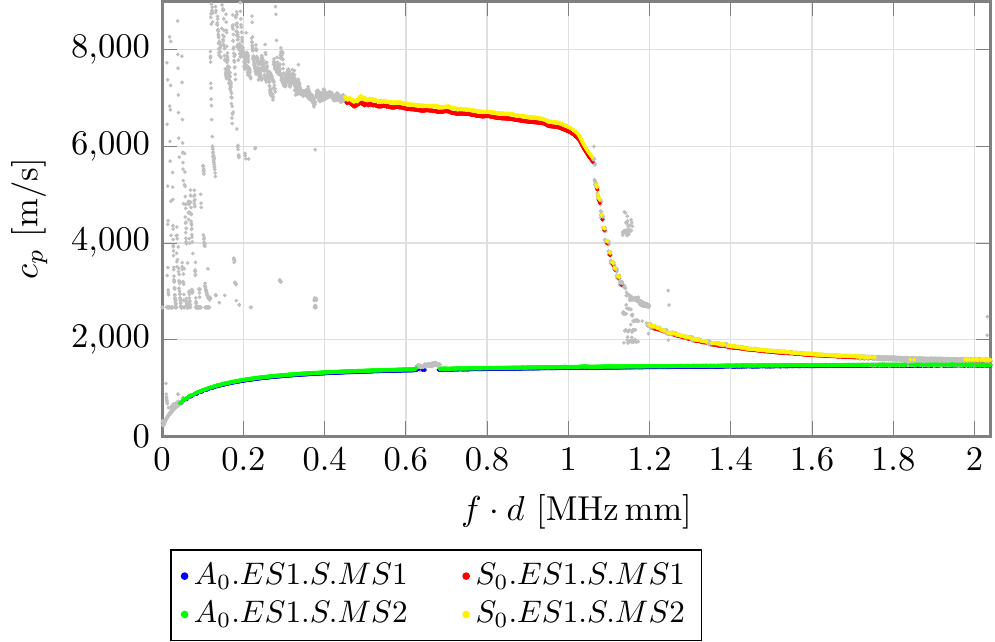}}
{\caption{Dispersion diagram for two measurements using the strip specimen with ES1.}
\label{fig:rep_strip_disp}}
\ffigbox[0.99\linewidth]
{\includegraphics[width=\linewidth]{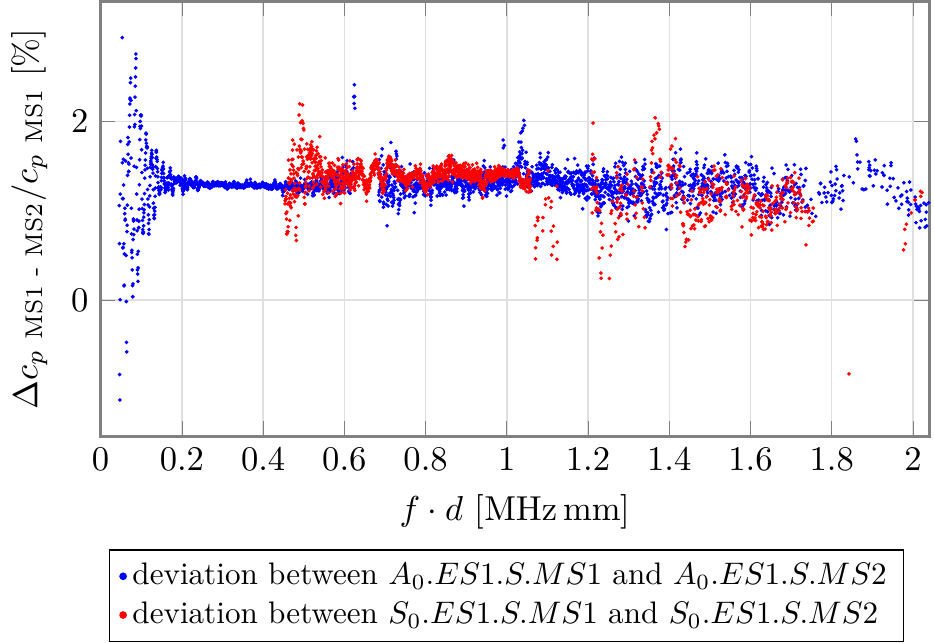}}
{\caption{Relative difference between two measurements using the strip specimen with ES1.}
\label{fig:rep_strip_diff}}
\end{floatrow}
\end{figure}

The reproducibility of the dispersion diagram results with ES1 is investigated using a strip specimen. The results of consecutive measurements are shown in Figure~\ref{fig:rep_strip_disp}. 
It is visible, that the $A_0$-mode can be evaluated over a wider frequency range than the $S_0$-mode. This is shown by the larger frequency range with evaluable data points compared to the $S_0$-mode. The reasons for the better detectability of the $A_0$-mode are mainly due to smaller wavelengths and a higher out-of-plane amplitude in the investigated frequency range. 
The minimally detectable wavelengths for the fundamental $A_0$- and $S_0$-mode, as described in Section~\ref{sec:data_sorting}, correspond to minimal frequency-thickness products of \SI{43}{\kilo \hertz \milli \meter} and \SI{443}{\kilo \hertz \milli \meter}, respectively.
\par
Figure~\ref{fig:rep_strip_diff} depicts a comparison of the two measurements by plotting the relative difference. The results for both fundamental modes show only slight deviations with a maximum error of approx. \SI{3}{\percent} and a mean relative difference of less than \SI{1,5}{\percent}. The larger scattering at the beginning of both detectable frequency ranges is the result of challenging measurement conditions in these frequency regions. As the wavenumbers become smaller with lower frequencies a larger measurement path would be needed for improved results. 
As stated above, the wavenumber of the two modes converges at higher frequencies. The low amplitudes of the $S_0$-mode interfere with the high amplitudes of the $A_0$-mode, resulting in a challenging wavenumber determination. 
\par
Considering the results shown here, it can be concluded that the consecutive measurements on a strip specimen with ES1 are in good agreement. 

\subsubsection{Comparison of plate specimen measurements with ES2}
\begin{figure}
\begin{floatrow}
\ffigbox[0.99\linewidth]
{\includegraphics[width=\linewidth]{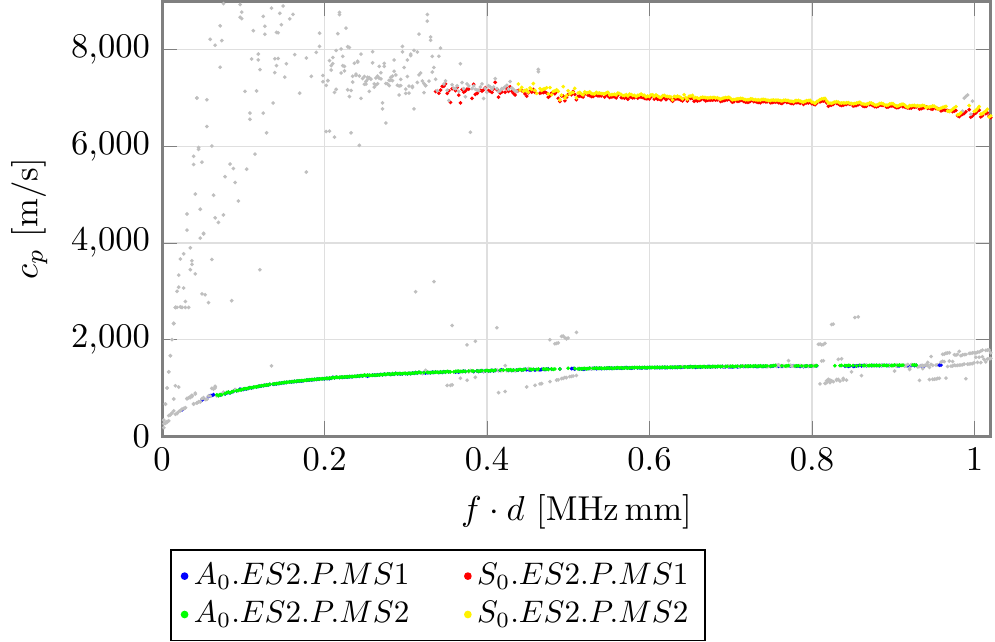}}
{\caption{Dispersion diagram for two measurements using the plate specimen with ES2.}
\label{fig:rep_plate_disp}}
\ffigbox[0.99\linewidth]
{\includegraphics[width=\linewidth]{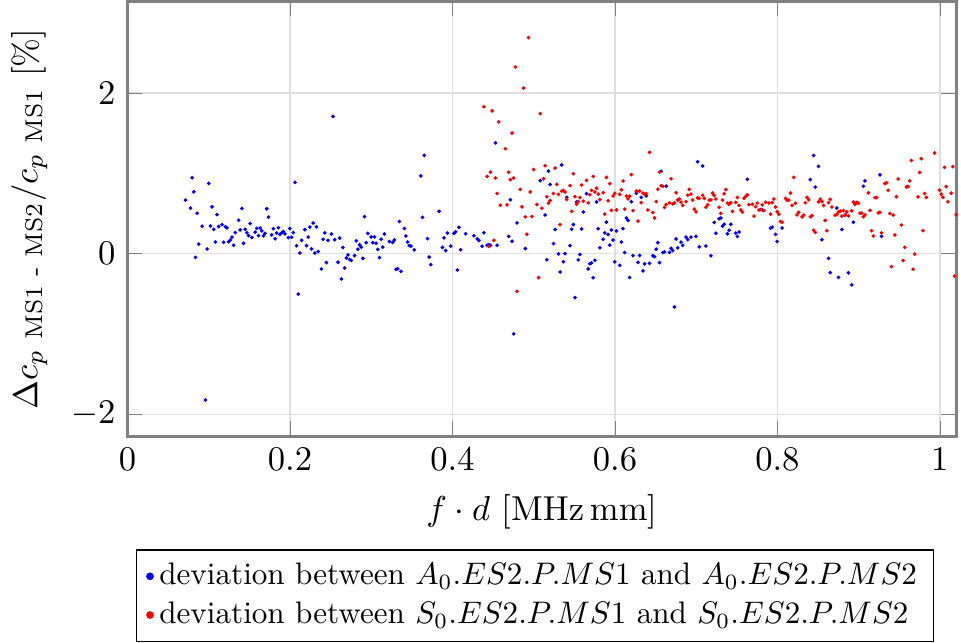}}
{\caption{Relative difference between two measurements using the plate specimen with ES2.}
\label{fig:rep_plate_diff}}
\end{floatrow}
\end{figure}

The results for the reproducibility investigation of ES2 are presented for a plate specimen. The same hardware is used to conduct the two different measurements. As for ES1, the overall course of the graphs for the two fundamental modes in Figure~\ref{fig:rep_plate_disp} are nearly the same for both measurements. In contrast to ES1, the maximum investigable frequency is limited due to the hardware, as described in Section~\ref{sec:data_sorting}.
The minimally detectable wavelengths correspond to minimal frequency-thickness products of \SI{26}{\kilo \hertz \milli \meter} and \SI{324}{\kilo \hertz \milli \meter} for the $A_0$- and $S_0$-mode, respectively.
\par
Figure~\ref{fig:rep_plate_diff} depicts the relative difference between the two measurements. The maximum error is approx. \SI{3}{\percent}, while the $A_0$-mode again shows slightly better reproducibility than the $S_0$-mode. The mean relative differences for the $A_0$- and $S_0$-mode are less than \SI{1}{\percent} and \SI{0,5}{\percent}, respectively.
\par
In general, ES2 is also suited to generate dispersion diagrams in FML repeatedly at high quality. In conclusion, a reliable generation of GUW dispersion diagrams in FML is shown with both setups. However, differences in performance between the two setups are to be expected, due to the different hardware equipment. This will be further discussed in the following section.

\subsection{Validating the measurement results with different experimental setups}\label{sec:reproducibility_ES}
Regarding data transfer and comparability in research cooperations, the question arises, whether the presented method is applicable with different experimental setups. The hypothesis is that the method can equally be applied to different setups, while the temporal and spatial resolutions, as well as measurement lengths and durations, define the investigable frequency-thickness product range.
\par
In Section~\ref{sec:reproducibility}, the reproducibility of the measurements was shown when using the same experimental setup. Now, the comparability of the measurements for an identical specimen is investigated by comparing the two measurement setups ES1 and ES2. For better comparability, the strip specimen is selected because of the same measurement path lengths, see Table~\ref{tab:specimen_geometries}. 
\par
\begin{figure}
\begin{floatrow}
\ffigbox[0.99\linewidth]
{\includegraphics[width=\linewidth]{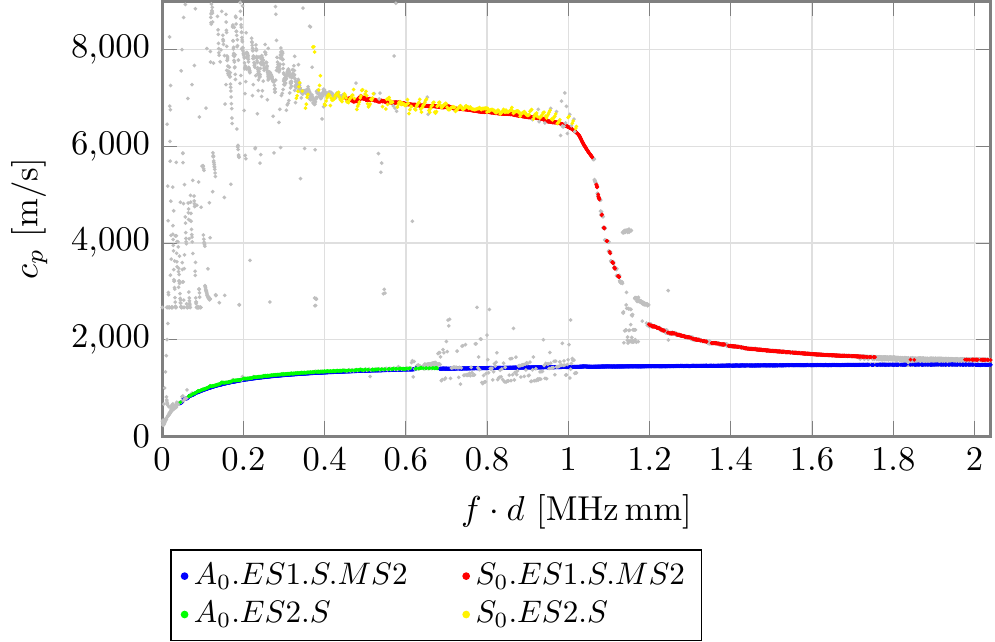}}
{\caption{Dispersion diagram of strip measurements with ES1 and ES2.}
\label{fig:setup_disp}}
\ffigbox[0.99\linewidth]
{\includegraphics[width=\linewidth]{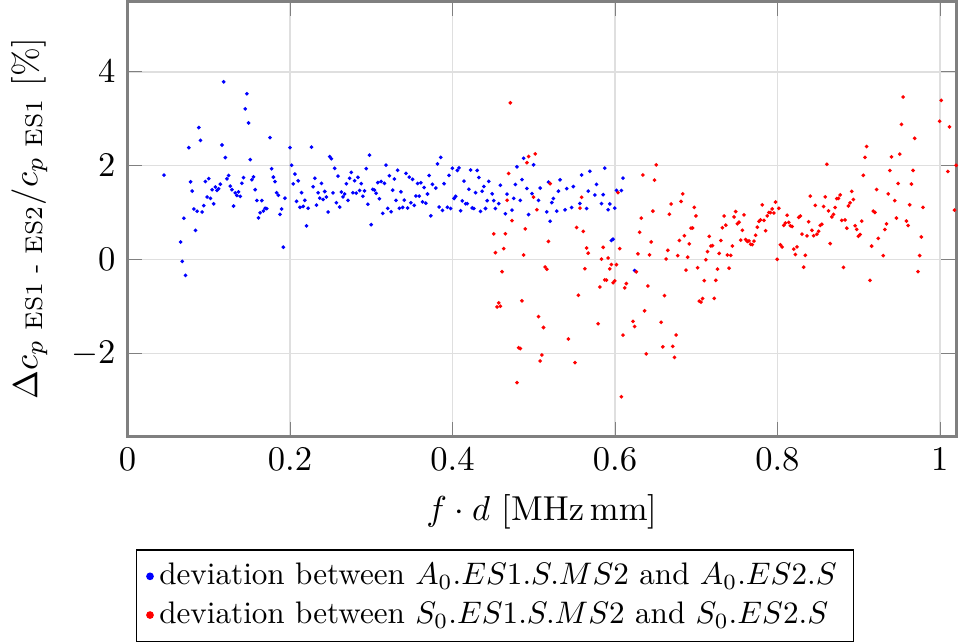}}
{\caption{Relative difference in strip measurements with ES1 and ES2.}
\label{fig:setup_diff}}
\end{floatrow}
\end{figure}

 
As mentioned in Section~\ref{sec:exp_setup}, the measurement setups are different with regard to the possible frequency range. Therefore, the comparison is restricted to the frequency limits in Table~\ref{tab:freq_ranges}. Besides this, the setups should produce comparable data with slightly clearer results for ES1 due to the overall qualitatively better measurement devices. This is visible from the smoothness of the data in Figure~\ref{fig:setup_disp}, where the effect is more pronounced in the $S_0$-mode due to the consistently lower measured wavenumbers. Otherwise, the graphic shows the expected picture, namely that both measurement setups can clearly show the course of the dispersive relationship. However, as expected, the detection of the $A_0$-mode yields significantly better results, which is partly due to the higher wavenumbers as well as the higher out-of-plane amplitudes in comparison to the $S_0$-mode. Figure~\ref{fig:setup_diff} provides a more detailed insight into the differences of the measurements by giving the percentage deviations of the measurements with respect to the measurement of ES1. It can be seen that despite of the different setups, only a small measurement deviation occurs, with a maximum of \SI{4}{\percent} and an average difference less than \SI{2}{\percent} between the setups. The better evaluability of the $A_0$-mode compared to the $S_0$-mode is thereby reflected in a significantly smaller scatter of the deviations. In conclusion, the measurements can be compared in FML even with different measurement setups, which enables data transfer in research cooperations.

\subsection{Validating the measurement results with different specimen sizes}
\label{sec:comparison_specimen_size}
In this subsection, measurements on strip and plate specimens are compared in order to investigate the influence of the specimens width. Generally speaking, specimens with large dimensions are always advantageous for the determination of GUW propagation phenomena, since reflections have less influence on the measurement~\cite{Rittmeier.2022}. The geometries of the strip and plate specimen under investigation are presented in Table~\ref{tab:specimen_geometries}. ES1 is chosen for comparison of the strip specimen to the reference plate since it covers a larger frequency range, see Section~\ref{sec:exp_setup}.\par
\begin{figure}
\begin{floatrow}
\ffigbox[0.99\linewidth]
{\includegraphics[width=\linewidth]{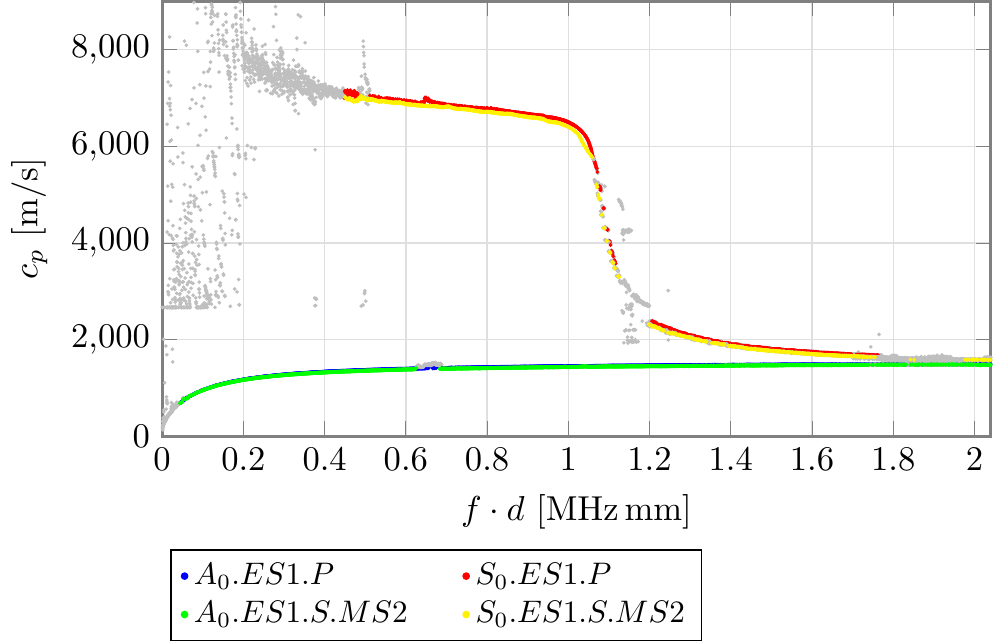}}
{\caption{Dispersion diagram of strip and plate measurements.}
\label{fig:spec_size_disp}}
\ffigbox[0.99\linewidth]
{\includegraphics[width=\linewidth]{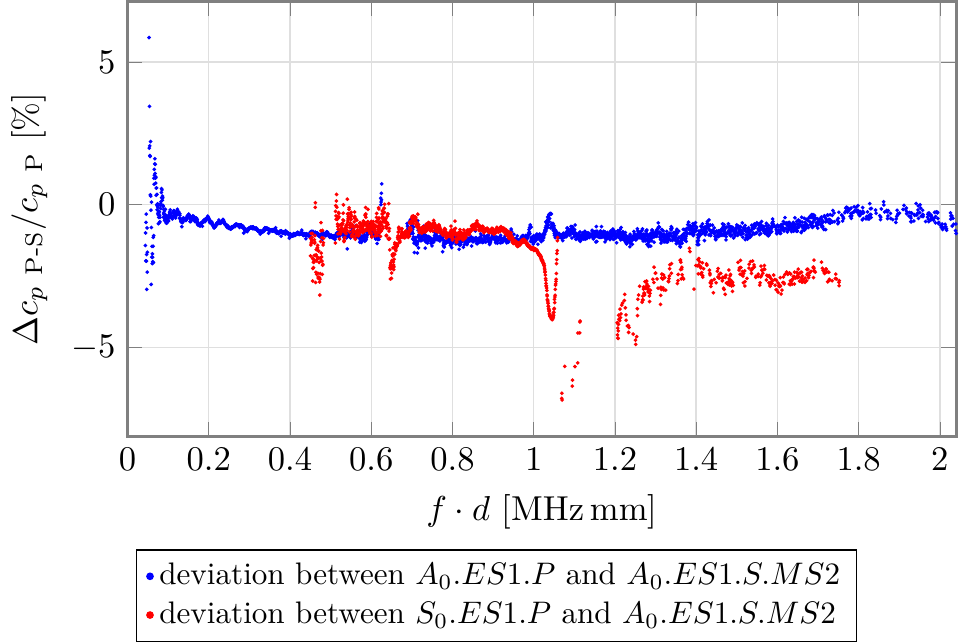}}
{\caption{Relative difference between strip (S) and plate (P) measurements.}
\label{fig:spec_size_diff}}
\end{floatrow}
\end{figure}

Figure~\ref{fig:spec_size_disp} shows the dispersion relation of the $A_0$- and $S_0$-mode for both measurements. Overall, no significant differences occur between the measurement in the plate and the strip specimen. The largest deviations occur in the frequency range of 1-\SI{1.2}{\mega \hertz \milli \meter} due to the crossing of the $SH$-modes as discussed in Section~\ref{sec:data_sorting}. Therefore, the larger deviations of the two measurements in this region cannot necessarily be attributed to the specimen geometry.
\par
This is supported by the difference plot in Figure~\ref{fig:spec_size_diff}. It can be seen, that the mean difference between the plate and strip specimens is below \SI{2}{\percent}, which is in the same range as the reproducibility of the measurement itself. Therefore, no significant difference between strip and plate specimen for the $A_0$-mode can be detected.
\par
As expected, there is more scattering in the relative difference for the $S_0$-mode. However, the absolute difference is less than \SI{5}{\percent} and for frequency-thickness products below \SI{1}{\mega \hertz \milli \meter}, the mean difference is below \SI{2}{\percent}. Although the $S_0$-mode difference is slightly higher compared to the previous measurements in this work, it is still considered reasonable.
\par
The use of a strip specimen instead of a large plate specimen is applicable for the dispersion diagram generation in this setup. The measurements shown indicate that for the considered frequency range the use of strip specimens with a width of \SI{110}{\milli \meter} instead of plates has only little effects on the results. Therefore, the use of strip specimens is very advantageous for cost reasons. It should be noted, that the statements made here about the usability of a strip specimen only refer to the here used specimen width and frequency region. However, the use of even narrower strips might increase the problem of edge reflections significantly.

\subsection{Comparison of experimental results and results from analytical framework}
\label{sec:comparison_analytic_experiment}
In this section, the experimental dispersion relations are compared to data from the numerical solution of the analytical framework in Section~\ref{sec:waves_in_laminate_structures}. For simplicity, this solution will be referred to as the analytical solution in the following. The measurements of the strip specimen in ES1 are used for the comparison. For the analytical solution two sets of material parameters, those from Hörberg~\cite{Horberg.2019} and Johnston~\cite{Johnston.1997}, are used to take into account the variations of these parameters from the literature.
\par
\begin{figure}
\begin{floatrow}
\ffigbox[0.99\linewidth]
{\includegraphics[width=\linewidth]{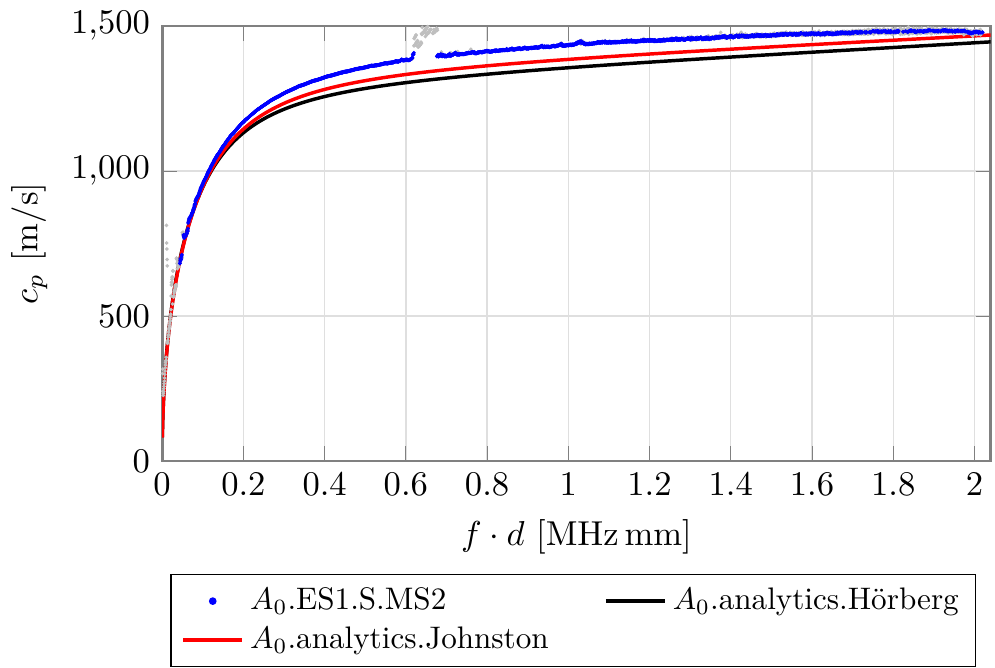}}
{\caption{Dispersion diagram of strip measurements and analytical solutions for the $A_0$-mode with different material parameters from literature~\cite{Horberg.2019,Johnston.1997}.}
\label{fig:an_disp_A0}}
\ffigbox[0.99\linewidth]
{\includegraphics[width=\linewidth]{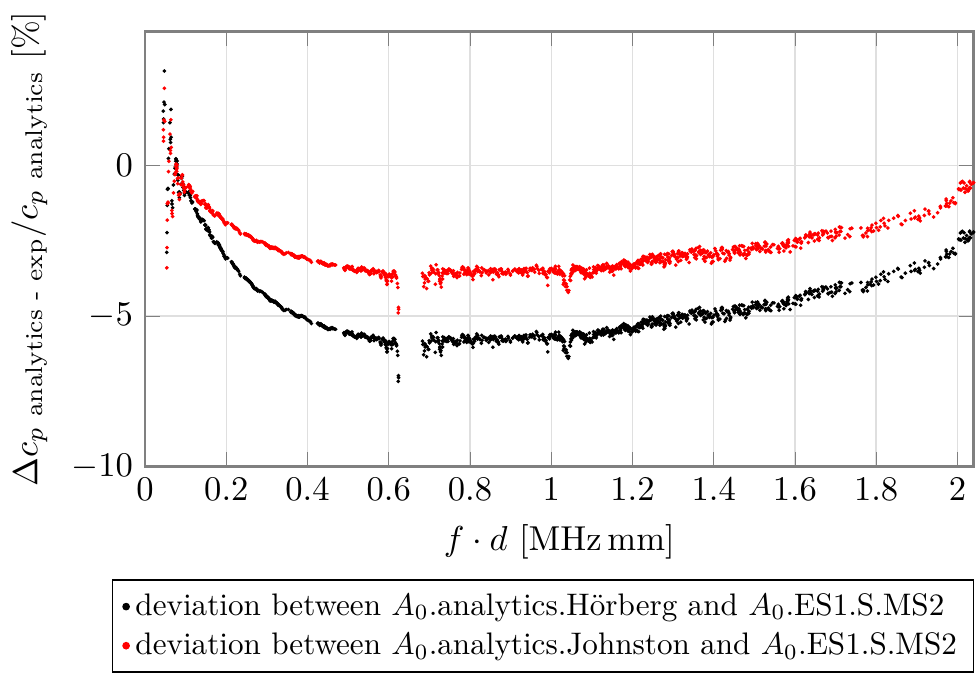}}
{\caption{Relative difference between strip measurements and analytical solution for the $A_0$-mode with different material parameters from literature~\cite{Horberg.2019,Johnston.1997}.}
\label{fig:an_diff_A0}}
\end{floatrow}
\end{figure}

In Figure~\ref{fig:an_disp_A0}, the analytical solutions are plotted against the measurement of the strip specimen in ES1 for the $A_0$-mode. The course of the measured data in the considered frequency range is in good agreement with the analytical values. Figure~\ref{fig:an_diff_A0} shows that the percentage deviation between experimental and analytical data is in the single-digit range for both analytical solutions. It can be stated that the parameters from~\cite{Johnston.1997} provide the more suitable solution, with deviations below approx. \SI{5}{\percent}. 
\par
\begin{figure}
\begin{floatrow}
\ffigbox[0.99\linewidth]
{\includegraphics[width=\linewidth]{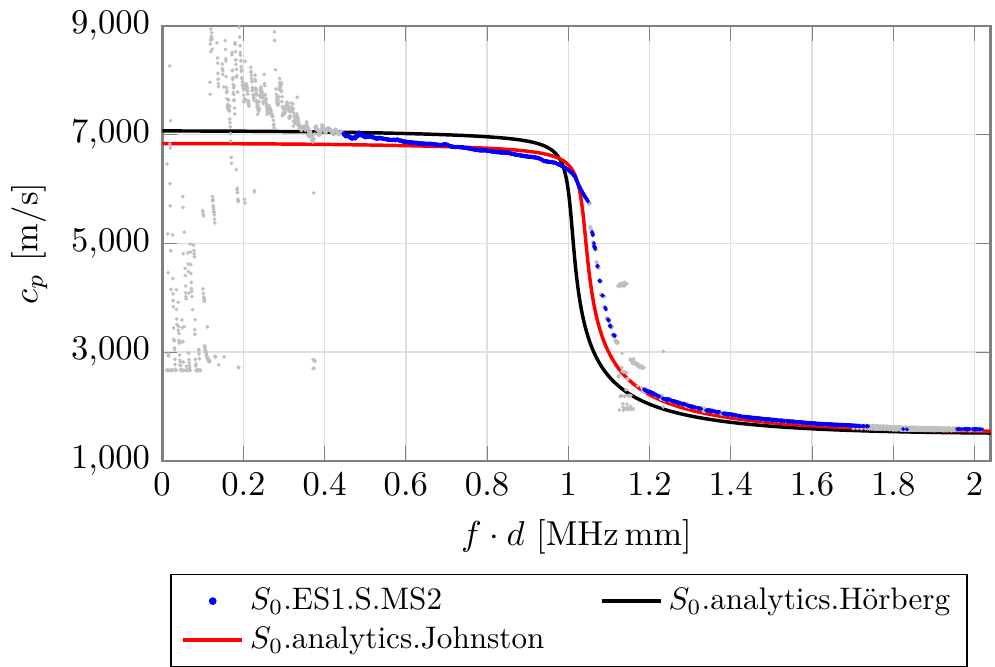}}
{\caption{Dispersion diagram of strip measurements and analytical solution for the $S_0$-mode with different material parameters from literature~\cite{Horberg.2019,Johnston.1997}.}
\label{fig:an_disp_S0}}
\ffigbox[0.99\linewidth]
{\includegraphics[width=\linewidth]{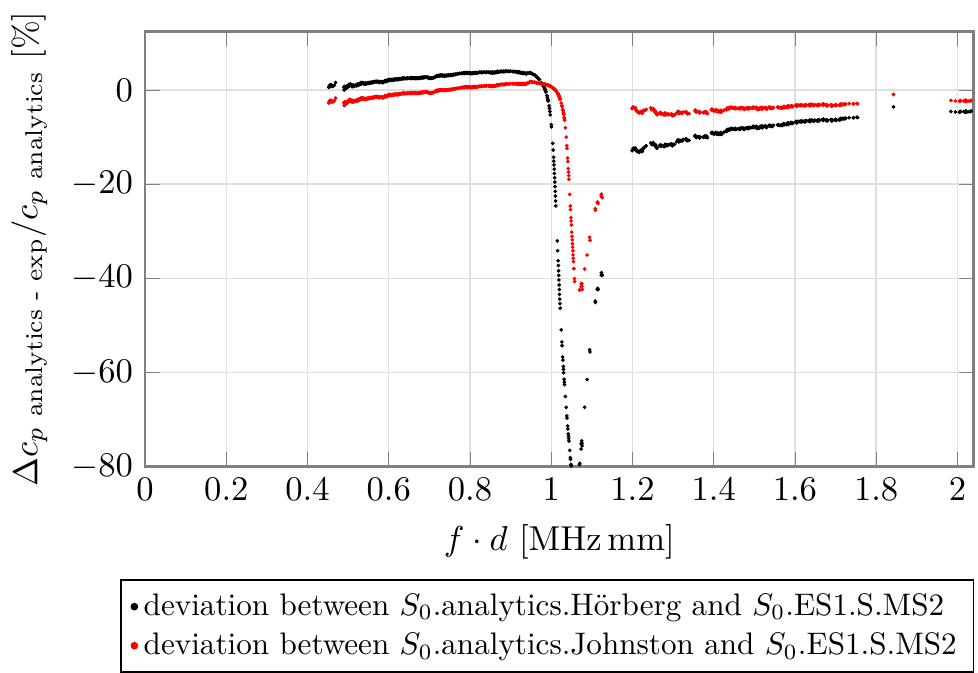}}
{\caption{Relative difference between strip measurements and analytical solution for the $S_0$-mode with different material parameters from literature~\cite{Horberg.2019,Johnston.1997}.}
\label{fig:an_diff_S0}}
\end{floatrow}
\end{figure}

In Figure~\ref{fig:an_disp_S0}, the same comparison is shown for the $S_0$-mode. Once again, the course of the measurement data is similar to that of the analytical solution. In contrast to the $A_0$-mode, larger differences can be found in the $S_0$-mode for certain frequency ranges. In the range from \SI{0.9}{\mega\hertz\milli\metre} to \SI{1.2}{\mega\hertz\milli\metre}, which contains the sharp decrease in phase velocity, a horizontal shift of this drop can be observed.
This is also clearly visible in Figure~\ref{fig:an_diff_S0}.
However, the high deviations are mostly due to the shift between the frequency ranges where the sharp decrease in phase velocity occurs. In general, there is good agreement up to approx. \SI{0.9}{\mega \hertz \milli \meter} and for frequency-thickness products higher than \SI{1.2}{\mega\hertz\milli\metre}. Again, the material data set from~\cite{Johnston.1997} shows better agreement to the experimental data compared to the material parameters found in~\cite{Horberg.2019}.
\par
The comparison of the measured data with the data from the analytical framework shows that the basic course of the dispersion diagrams can be represented properly. In addition, it can be assumed that an adjustment of the material parameters within a reasonable range would lead to a further improved agreement between the analytical and the experimental data. The investigation of the individual material parameters required for this purpose will be considered in depth in the further course of this research.
\par
Another aspect that was not considered so far is the inherent manufacturing-induced residual stress state in the FML. Interlaminar stresses develop in the CFRP-steel specimen, mainly due to the difference in the coefficients of thermal expansion between steel and CFRP, in combination with temperatures of up to \SI{180}{\degreeCelsius} during the manufacturing process. This stress state results in inherent tensile stresses in the metal layers and compressive stresses in the fiber layers after curing. For the layup at hand, the metal layers are subjected to significant tensile stresses that account for up to \SI{20}{\percent} of its tensile strength~\cite{Wiedemann.2022,Wiedemann.2022b}. The residual stresses in the CFRP layers, on the other hand, suggest that parameters in the compression range would have to be taken into account for this material. These parameters can deviate from those in the tensile range, see~\cite{NCAMP.2011}. However, this internal stress state is not yet accounted for in the analytical calculation and might be an additional reason for the deviations between analytical and experimental data.

\section{Conclusion}
This work extends the use of 2d Fourier transformations for the experimental determination of dispersion relations of GUWs for isotropic materials, to the use in anisotropic and inhomogeneous FML structures made out of CFRP and steel. Therefore, a method based on custom multi-frequency excitation signals by Barth et al.~\cite{Barth.ExpM.2022} has been used to experimentally determine dispersion relationships of GUWs in FML over large frequency ranges. Comparable to the use in isotropic materials, a very high accuracy could be achieved in FML that exhibit a strong anisotropic behavior. The results were successfully reproduced with repeated measurements, different measurement setups as well as different specimen geometries.
\par
A comparison between the experimental data and data from numerical solutions of the analytical framework show an excellent agreement. Small differences can be explained by deviations and uncertainties in the material parameters used. Explanatory approaches for the deviations are pointed out and will be addressed in future work. Nevertheless, the results provide a profound basis for the evaluation of GUW based SHM in FML.
\par
In the further course of the research, the measurements are to be extended and supplemented by modeling using FEM. The investigations will deal with the influences of manufacturing-related residual stress states in FML, to provide fundamental understanding of the influence of theses stress states on the wave propagation of GUWs.









\section*{Acknowledgment}
The authors expressly acknowledge the financial support for the research work on this article within the Research Unit~3022 “Ultrasonic Monitoring of Fibre Metal Laminates Using Integrated Sensors” by the German Research Foundation (Deutsche Forschungsgemeinschaft~(DFG)).

\printcredits

\section*{Data availability}
The publication of the data that support the findings of this study is in progress and will be openly available at the time this study will be published.

\bibliographystyle{elsarticle-num-names.bst}

\bibliography{Literature.bib}



\end{document}